\documentclass[aps,twocolumn,superscriptaddress,groupedaddress,preprintnumbers,nofootinbib,twoside]{revtex4-1} 

\usepackage{graphicx}  
\usepackage{dcolumn}   
\usepackage{bm}        
\usepackage{amssymb}   
\usepackage[utf8]{inputenc}
\usepackage{amsmath}
\usepackage{amssymb}
\usepackage{amsthm}
\usepackage{color}
\usepackage{rotating}
\usepackage{graphicx}
\usepackage{float}
\usepackage{wrapfig}
\usepackage{enumitem}
\usepackage{listings}
\usepackage{caption}
\usepackage{wrapfig}
\usepackage{array}
\usepackage{wasysym}
\usepackage{braket}
\usepackage[titletoc]{appendix}
\usepackage{tikz}
\usetikzlibrary{calc}
\usepackage{makeidx}
\usepackage{slashed}
\usepackage[margin={2cm,2cm}]{geometry}

\usepackage{subcaption}
\usepackage{natbib}

\begin{document}

\title{	Investigating Dark Matter and MOND Models with Galactic Rotation Curve Data\\
	\small{Analysing the Gas-Dominated Galaxies}}

\author{Jonas Petersen}\email{petersen@cp3.sdu.dk}\affiliation{$CP^3$-Origins, University of Southern Denmark, Campusvej 55, DK-5230 Odense M, Denmark}
\date{\today}

\preprint{Preprint: CP$^3$-Origins-2019-12 DNRF90}

\begin{abstract}
In this study the geometry of gas dominated galaxies in the SPARC database is analyzed in a normalized $(g_{bar},g_{obs})$-space ($g2$-space), where $g_{obs}$ is the observed centripetal acceleration and $g_{bar}$ is the centripetal acceleration as obtained from the observed baryonic matter via Newtonian dynamics. The normalization of $g2$-space significantly reduce the effect of both random and systematic uncertainties as well as enable a comparison of the geometries of different galaxies. Analyzing the gas-dominated galaxies (as opposed to other galaxies) further suppress the impact of the mass to light ratios.\newline
It is found that the overall geometry of the gas dominated galaxies in SPARC is consistent with a rightward curving geometry in the normalized $g2$-space (characterized by $r_{obs}>r_{bar}$, where $r_{bar}=\arg \max_r[g_{bar}(r)]$ and $r_{obs}=\arg \max_r[g_{obs}(r)]$). This is in contrast to the overall geometry of all galaxies in SPARC which best approximates a geometry curing nowhere in normalized $g2$-space (characterized by $r_{obs}=r_{bar}$) with a slight inclination toward a rightward curving geometry. The geometry of the gas dominated galaxies not only indicate the true (independent of mass to light ratios to leading order) geometry of data in $g2$-space (which can be used to infer properties on the solution to the missing mass problem) but also - when compared to the geometry of all galaxies - indicate the underlying radial dependence of the disk mass to light ratio. 

\end{abstract}

\maketitle

\section{Introduction}
The missing mass problem is established by a host of astronomical observations, including galaxy rotation curve measurements~\citep{Rubin:1980zd,Albada,Bosma}, studies of the dynamical behaviour of galaxy clusters~\citep{Zwicky:1933gu,Clowe:2006eq} and measurements of the cosmic microwave background~\citep{Ade:2015xua}. Solutions has been proposed either in terms of some unobserved matter (dark matter), as a modification of classical dynamics or a combination of the two. Particle dark matter has been argued to be in overall agreement with observations on all the aforementioned scales, however there remain several significant challenges on galactic scales~\citep{Bullock:2017xww,Flores:1994gz,Moore:1999nt,BoylanKolchin:2011de}.\newline 
The idea of modified classical dynamics is primarily motivated by studies on galactic scales which reveal a close relationship between the baryonic matter in rotationally supported galaxies and the observed dynamical behaviour of galaxies in regions dominated by the solution to the missing mass problem~\citep{Tully:1977fu,Faber1976,McGaugh:2000sr,McGaugh:2016leg,Lelli:2017vgz}. These observations inspired modified Newtonian dynamics (MOND) in 1983~\citep{Milgrom:1983ca}, which come in the form of theories of modified inertia or gravity. Both MOND classes have struggled on larger scales for example in accounting for the entire gravitational anomaly in galaxy clusters~\citep{Sanders:2002ue} and cosmological observations~\citep{Skordis:2005xk,Dodelson:2006zt,Dodelson:2011qv}. Despite this, the MOND theories remain - at the very least - an inspiration to both dark matter and modified gravity model builders.\newline
The discussion surrounding MOND was reinvigorated when~\citep{McGaugh:2016leg,Lelli:2017vgz} discovered that galactic rotation curve data from the SPARC database~\citep{Lelli:2016zqa} globally follow an analytical relation (dubbed the radial acceleration relation (RAR)) which carry the unique characteristics of MOND modified inertia. The result of~\citep{McGaugh:2016leg,Lelli:2017vgz} even inspired dark matter model builders to reproduce the RAR, with the unique characteristics of MOND modified inertia, within the dark matter hypothesis e.g. ~\citep{Chashchina:2016wle,Edmonds:2017zhg,Dai:2017unr,Cai:2017buj,Berezhiani:2017tth}. In 2017~\citep{Petersen:2017klw} performed a local analysis of the RAR in which the function was fit to large radii and subsequently compared to data at small radii. In this study it was found that data at small radii deviate from the RAR fitted to large radii with between $3.5\sigma$ and $>8\sigma$ confidence depending on technical details. This result show that small radii deviate significantly from large radii and in so doing the predictions of the RAR and MOND modified inertia. \citep{Li:2018tdo} subsequently found that the RAR is in overall agreement with individual galaxy fits. This is consistent with the results in \citep{Petersen:2017klw}, since there only the discrepancy between large and small radii was investigated (small radii constitute only around $\sim 5\%$ of points in the SPARC database). The discussion continued with \citep{Frandsen:2018ftj} finding a tension of more than $8\sigma$ between a particular prediction of MOND modified inertia in general (independent of interpolation function) and data from the SPARC database. \citep{Frandsen:2018ftj} also analyse the geometry of data relative to theoretical predictions in $(g_{bar},g_{obs})$-space ($g2$-space), where $g_{obs}$ is the observed centripetal acceleration and $g_{bar}$ is the centripetal acceleration as obtained from the observed baryonic matter via Newtonian dynamics. Studying the $g2$-space geometries of data from different galaxies can provide information about the details of the solution to the missing mass problem and possibly reject and shape models of both dark matter and modified gravity in the future. \citep{Frandsen:2018ftj} find that geometries from different galaxies can be categorized into three sub-categories - here defined as rightward ($r_{bar}<r_{obs}$, characteristic of e.g. pseudo-isothermal dark matter and MOND modified gravity), leftward ($r_{bar}>r_{obs}$ characteristic of e.g. NFW dark matter) and nowhere ($r_{bar}=r_{obs}$, characteristic of e.g. MOND modified inertia) - according to the relative magnitude of $r_{bar}$ and $r_{obs}$, where $r_{bar}$ is the radius corresponding to the maximum in $g_{bar}$ and $r_{obs}$ is the radius corresponding to the maximum in $g_{obs}$. \citep{Frandsen:2018ftj} show that all three subcategories appear to be present in the data and discuss how such knowledge can be used to infer properties of the solution to the gravitational anomaly.\newline

The distribution of galaxies across the different sub-categories of geometries found by~\citep{Frandsen:2018ftj} rely on the assumption that the mass to light ratios of the disk and bulge do not have a (significant) radial dependency. In this article this assumption will be investigated and the distribution of $g2$-space geometries across the different sub-categories will be analysed further. This will be done by analysing the gas-dominated galaxies of the SPARC database. Gas-dominated galaxies do not, to leading order, depend on the mass to light ratios of the disk and bulge\footnote{It turns out that the gas-dominated galaxies in the SPARC database do not have bulges, so in this article, only the radial dependence of the mass to light ratio of the galactic disk is investigated.}, and as such comparing the $g2$-space geometries of gas-dominated galaxies to those of all galaxies will provide information about a potential radial dependence of the mass to light ratios as well as indicating the true distribution of $g2$-space geometries across sub-categories  (rightward, leftward and nowhere).
 
\section{Data analysis}
The SPARC database consists of rotation curve data from $175$ rotationally supported galaxies \citep{Lelli:2017vgz,Lelli:2016zqa}. The database provide observed rotational velocities ($v_{obs}$), along with the associated uncertainties ($\delta v_{obs}$), as well as distance ($D$) and inclination ($\alpha$) measurements for each galaxy and the rotational velocities due to the baryonic components ($v_{gas}\equiv v_g,v_{bul}\equiv v_b$ and $v_{disk}\equiv v_d$). In line with \citep{McGaugh:2016leg,Lelli:2017vgz} a $10\%$ relative uncertainty on the mass of the gas, $m^g_i$, ($\delta m^g_i$) is taken.\newline\newline
Following \citep{Lelli:2017vgz} $22$ galaxies are discarded from the analysis based on the inclination angle of the galaxy and a further quality criteria defined in \citep{Lelli:2017vgz}. Following \citep{Frandsen:2018ftj,Petersen:2017klw} $1$ additional galaxy is discarded due to large negative values of $v_g$ at the innermost radii, leaving $152$ galaxies making up $3143$ data points in $g2$-space. \newline\newline

\noindent The baryonic velocity ($v_{bar}$) for the $k$'th data point in the $i$'th galaxy is computed via 
	\begin{equation}
		v_{bar}(r_{k,i})=\sqrt{|v_{g}(r_{k,i})|v_{g}(r_{k,i})+\tilde{\Upsilon}^{d}_iv_{d}^2(r_{k,i})+\tilde{\Upsilon}^{b}_iv_{b}^2(r_{k,i})},
		\label{v_bar}
	\end{equation}
where $\tilde{\Upsilon}^{d}_i$ and $\tilde{\Upsilon}^{b}_i$ are the unitless mass to light ratios of the $i$'th galaxy, it has been used that $v_d,v_b>0$ and the radial measurement corresponding to the individual measurements of $v_g,v_d$ and $v_b$ is used to label the individual data points. In line with \citep{Mcagugh:2014,Lelli:2016zqa} $\tilde{\Upsilon}^{d}_i=0.5$ will be taken with a $25 \%$ uncertainty. 

\subsection{Normalized $g2$-space}
Following \citep{Frandsen:2018ftj} the $g2$-space coordinates are normalized according to the data point corresponding to $r_{bar}$. Doing so eliminates the systematic uncertainty from the galaxy distance and inclination angle as well as reduces the systematic uncertainty from the mass to light ratios\footnote{The effect of the mass to light ratios is already reduced by considering galaxies which have dynamics dominated by gas. The effect is further reduced by constructing normalized coordinates.} and gas measurements. 
In order to reduce the systematic uncertainty introduced by the normalization of the coordinates as well as reducing the vulnerability to taking a statistical outlier as normalization point, the normalization is taken to be the average between the point corresponding to $r_{bar}$ and the two adjacent points, i.e. $r_{bar\pm 1}$. Hence, the normalized coordinates are given by
\begin{equation}
\begin{split}
&\hat{g}_{obs}(r_{k,i},r_{bar\pm 1,i})=\frac{g_{obs}(r_{k,i})}{\frac{1}{N_i}\sum _{s= bar\pm 1}g_{obs}(r_{s,i})}\\
&\qquad\qquad\qquad\quad\,\,\,\equiv \frac{g_{obs}(r_{k,i})}{\braket{g_{obs}(r_{bar,i})}},\\\\
\end{split}
\label{norm22}
\end{equation}
\begin{equation}
\begin{split}
&\hat{g}_{bar}(r_{k,i}, r_{bar\pm 1,i})=\frac{g_{bar}(r_{k,i})}{\frac{1}{N_i}\sum _{s= bar\pm 1}g_{bar}(r_{s,i})}\\
&\qquad\qquad\qquad\quad\,\,\,\,\equiv \frac{g_{bar}(r_{k,i})}{\braket{g_{bar}(r_{bar,i})}},\\
\end{split}
\label{norm2}
\end{equation}
where $N_i$ denote the number of points $s$ span over. If $r_{bar}$ is the point at lowest radius, only $r_{bar}$ and $r_{bar+1}$ are included in the sum over $s$. Similarly if $r_{bar}$ is the point at largest radius only $r_{bar}$ and $r_{bar-1}$ are included in the sum over $s$. In these cases $N_i=2$, otherwise $N_i=3$.

\subsection{Gas-dominated galaxies}
\begin{figure*}[ht]
	\centering
	\captionsetup{width=1\textwidth}
	\begin{subfigure}{0.3\textwidth}
		\includegraphics[width=1\textwidth]{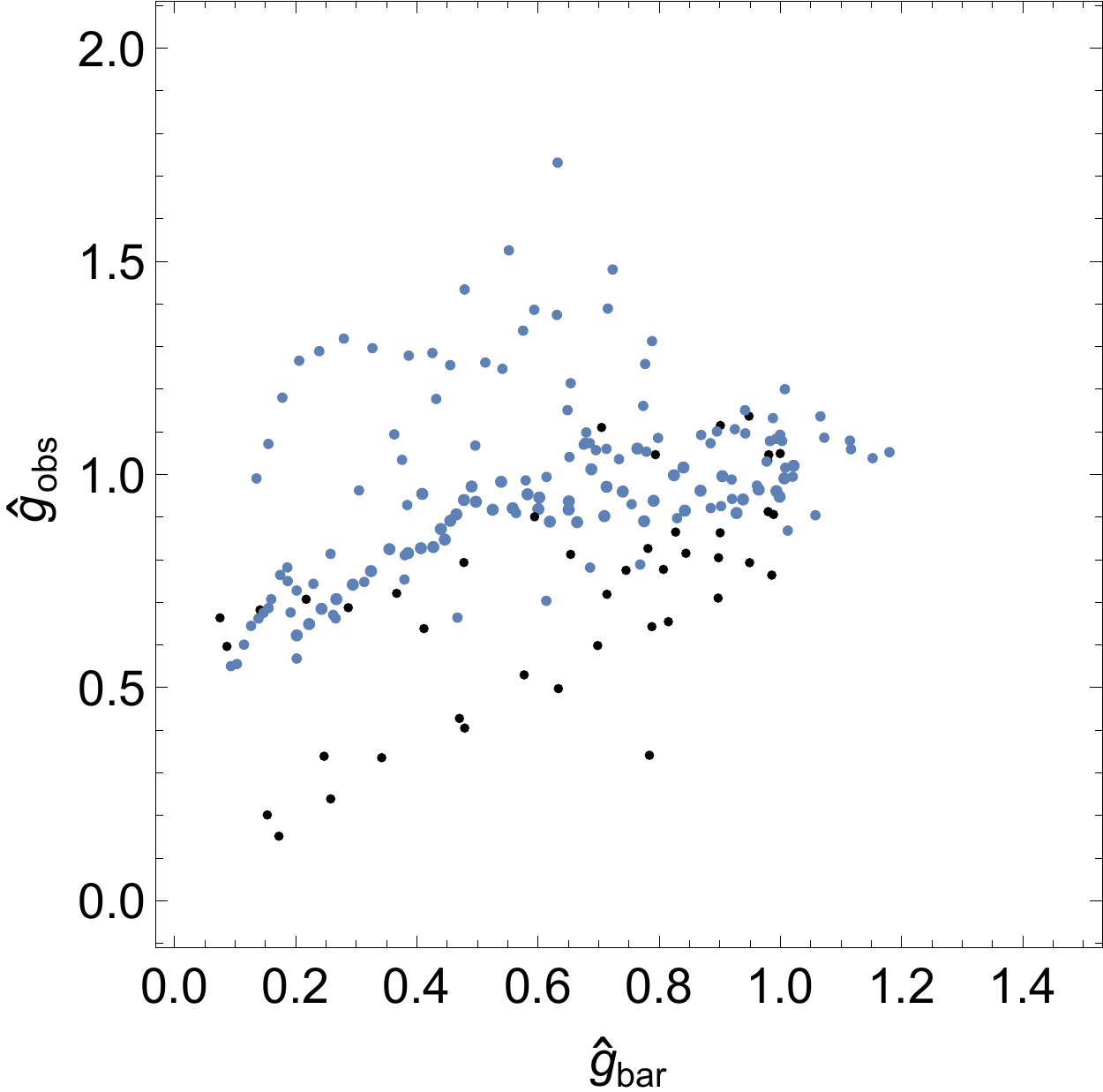}
	\end{subfigure}
	\begin{subfigure}{0.3\textwidth}
		\includegraphics[width=1\textwidth]{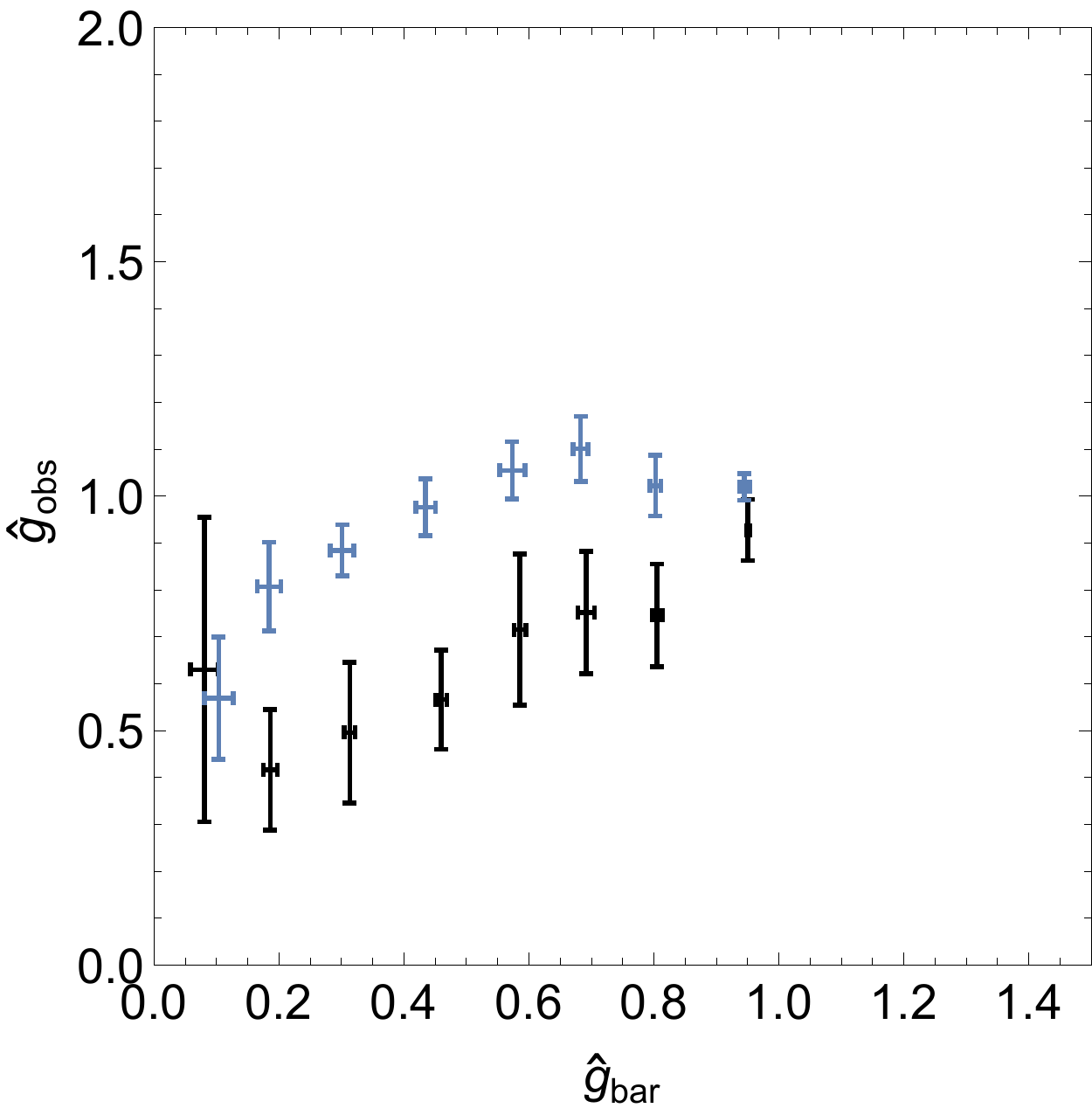}
	\end{subfigure}\\
	\begin{subfigure}{0.3\textwidth}
		\includegraphics[width=1\textwidth]{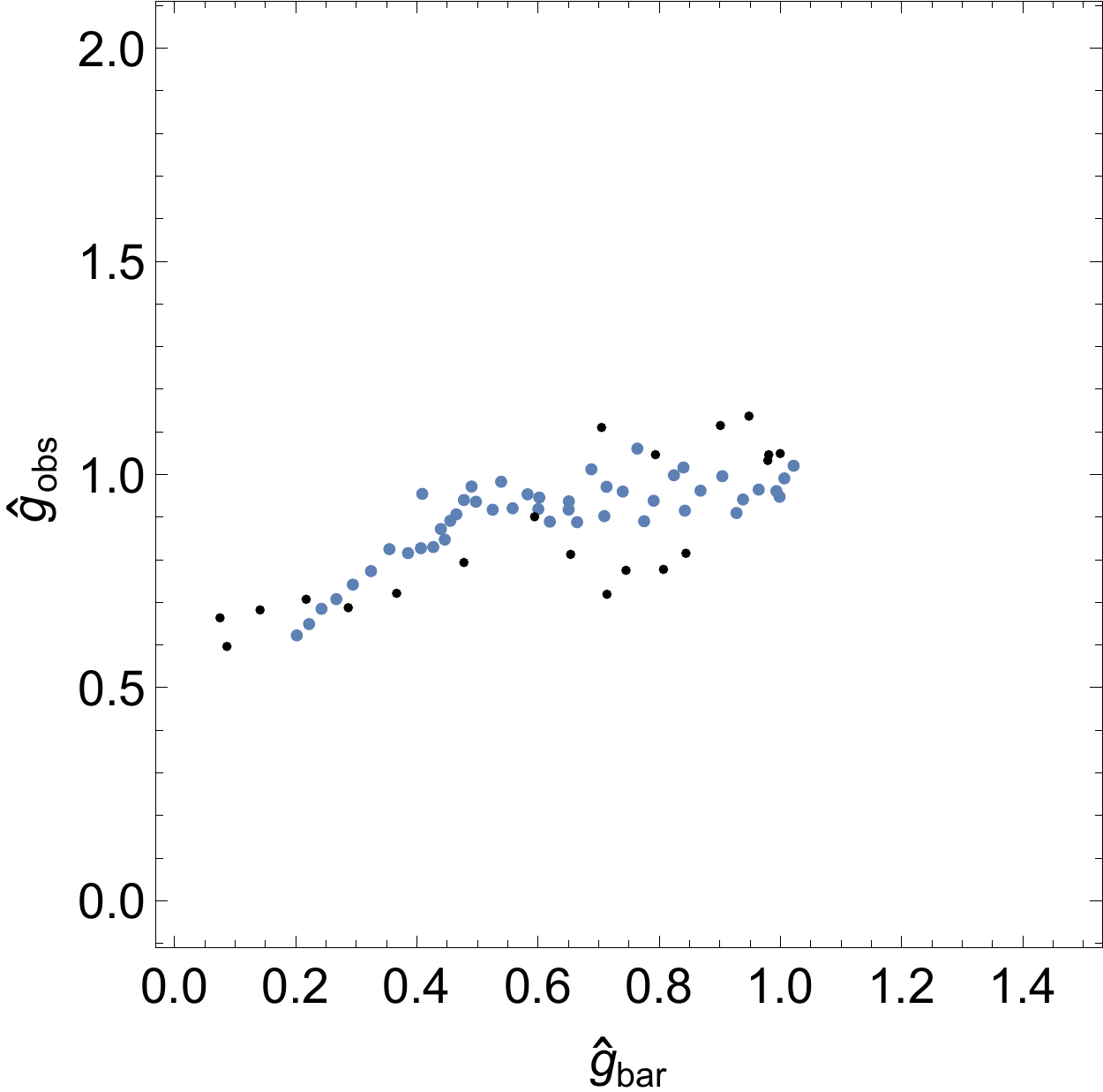}
	\end{subfigure}
	\begin{subfigure}{0.3\textwidth}
		\includegraphics[width=1\textwidth]{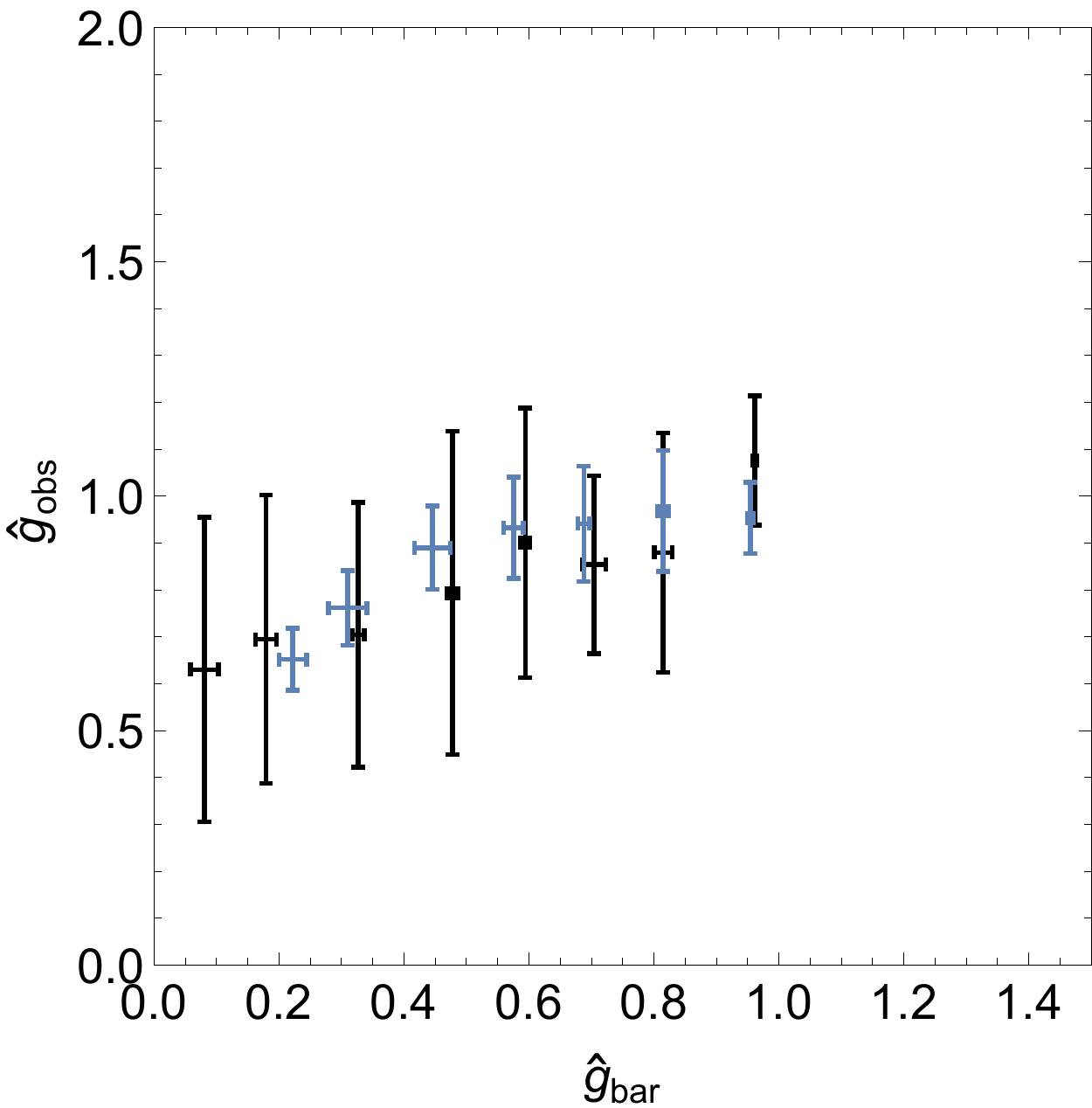}
	\end{subfigure}\\
	\begin{subfigure}{0.3\textwidth}
		\includegraphics[width=1\textwidth]{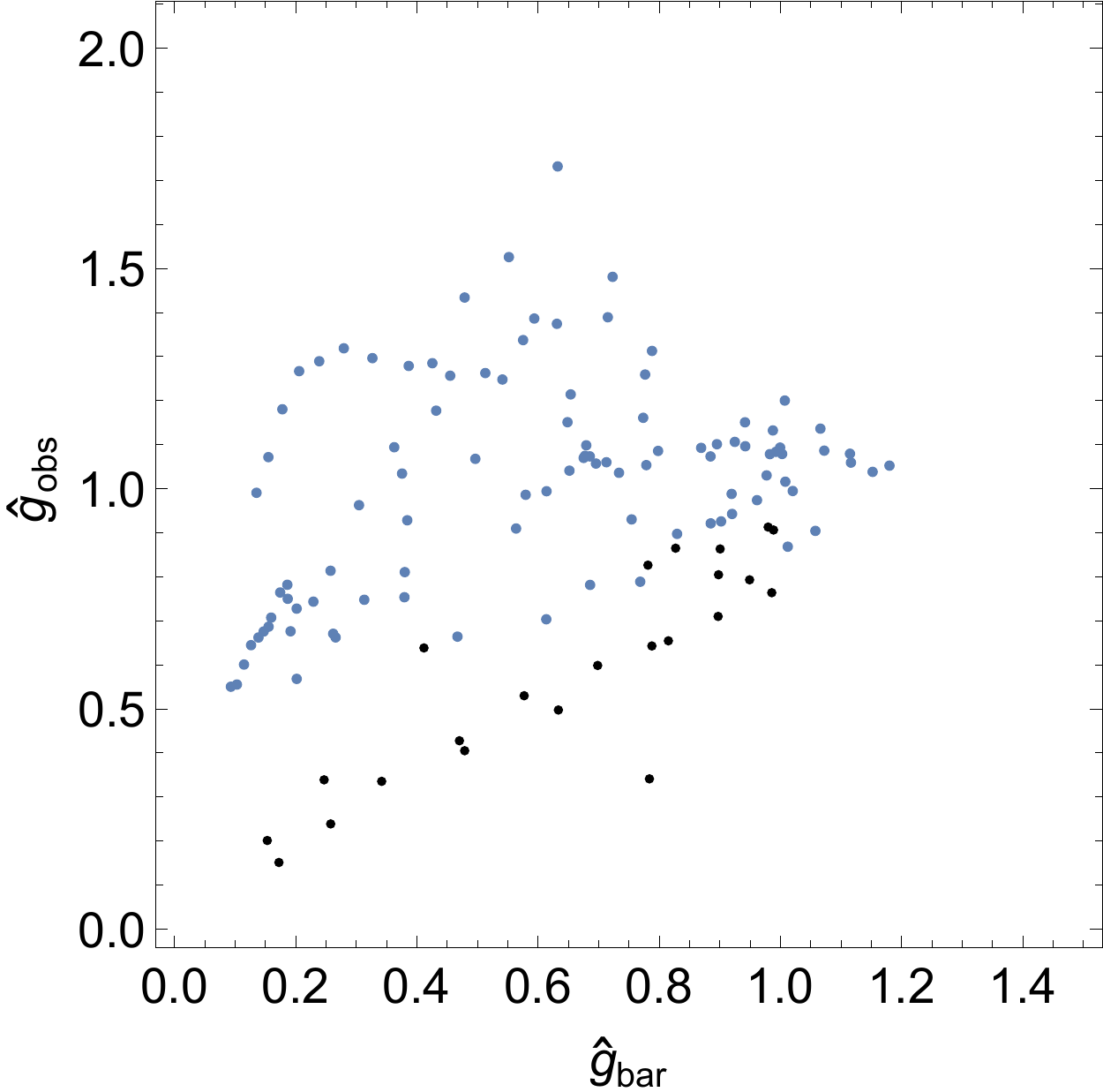}
	\end{subfigure}
	\begin{subfigure}{0.3\textwidth}
		\includegraphics[width=1\textwidth]{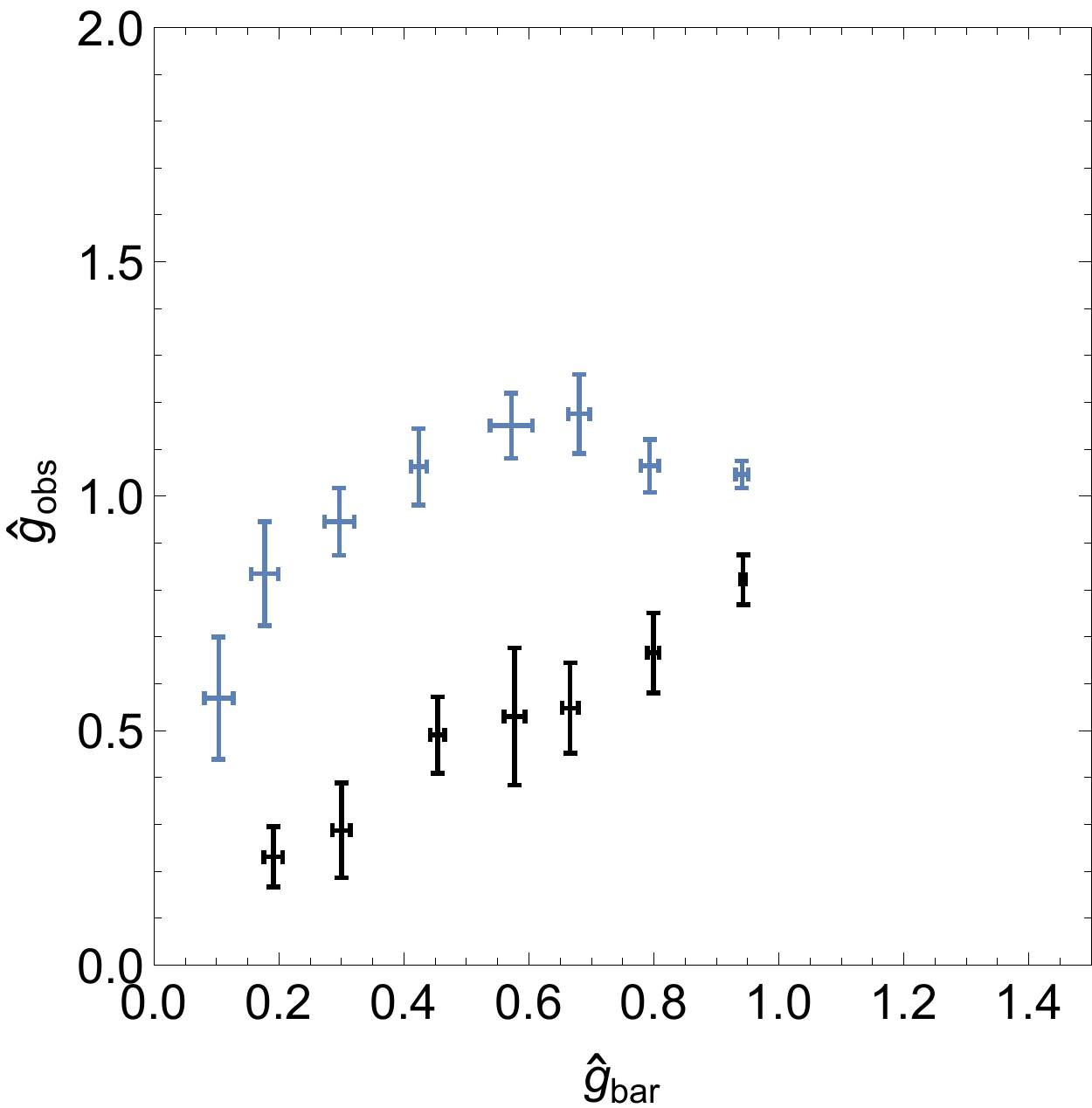}
	\end{subfigure}
	\caption{Left column: Data from the $11$ gas-dominated galaxies from the SPARC database in $g2$-space normalized according to equations \eqref{norm2} and \eqref{norm22}. Black points are at $r<r_{bar}$, whereas blue points are at $r\geq r_{bar}$. Note that the uncertainty of individual points are not shown in order to not clutter the plot. Top left panel: data from all $11$ galaxies. Middle left panel: data from leftward galaxies. Bottom left panel: data from rightward galaxies. Right column: Binned plots of the left column.}
	\label{fig:1}
\end{figure*}

\begin{figure*}[!htb]
	\centering
	\captionsetup{width=1\textwidth}
	\begin{subfigure}{0.3\textwidth}
		\includegraphics[width=1\textwidth]{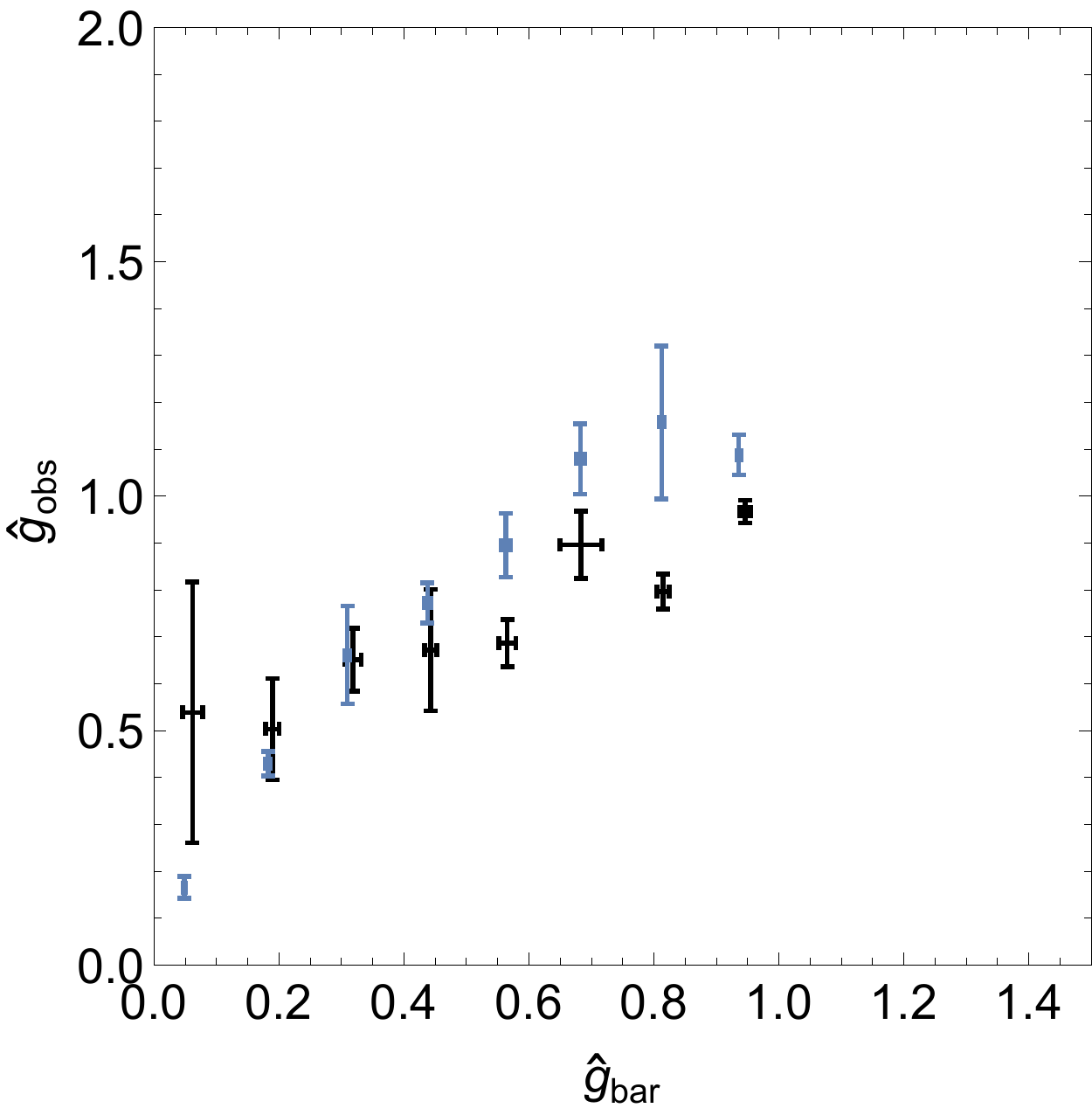}
	\end{subfigure}
	\begin{subfigure}{0.3\textwidth}
		\includegraphics[width=1\textwidth]{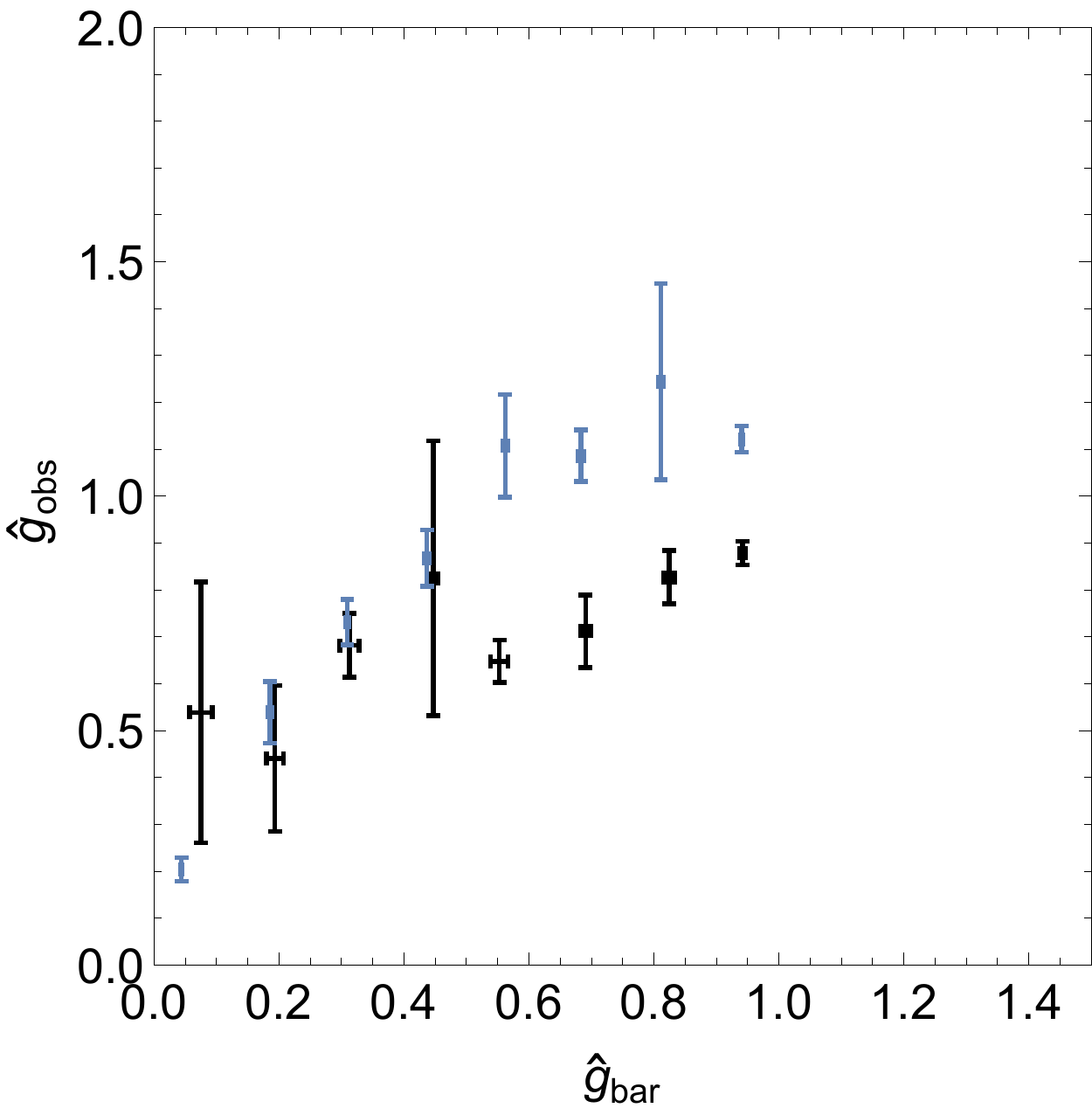}
	\end{subfigure}
	\caption
	{Data from the $152$ galaxies in the SPARC database binned according to equation \eqref{norm2} and \eqref{norm22}. Left panel: Adopt $\tilde{\Upsilon}^{d}_i=0.5$ and $\tilde{\Upsilon}^{b}_i=0.7$ with $25 \%$ uncertainty. Right panel: Adopt $\tilde{\Upsilon}^{d}_i=0.5e^{-(\frac{5}{4}(0.3)^3+0.13)(\frac{r}{r_d}-1)}$ \citep{Portinari:2009ap} and $\tilde{\Upsilon}^{b}_i=0.7$ with $25 \%$ uncertainty.}
	\label{fig:5}
\end{figure*}

A gas-dominated galaxy is conventionally defined in terms of the fraction of the galactic baryonic mass that is comprised of gas. In relation to obtaining data (in $g2$-space) independent of the stellar matter in the galaxies over the largest possible radial range, it is however more expedient to define a gas dominated galaxy in terms of the fraction of data points for which the gas-velocity is the largest baryonic velocity component. Hence, if
\begin{equation}
v_g(r_{k,i})>\sqrt{v_d^2(r_k,i)\tilde{\Upsilon}^{d}_i}\, \wedge \, v_g(r_{k,i})>\sqrt{v_b^2(r_k,i)\tilde{\Upsilon}^{b}_i}
\label{gdom}
\end{equation}
the $k$'th data point of the $i$'th galaxy is registered as gas-dominated. If more than $80\%$ of data points from a given galaxy is gas-dominated, the galaxy is classified as gas-dominated. This definition classifies 11/152 galaxies, defined as the set $G$ (189 data points in $g2$-space), of the SPARC database as gas-dominated;
\begin{equation}
\begin{split}
G=\{&UGCA444, UGC12732, UGC11820, UGC04483\\
&, UGC00731, NGC3741, NGC3109\\
&,F583 - 1, DDO168, DDO161, DDO154\}.
\end{split}
\end{equation}
Appendix \ref{app:plots} shows rotation curve plots of the galaxies in $G$. 9/11 galaxies are classified as rightward ($r_{obs}>r_{bar}$) whereas the remaining $2$ are classified as leftward ($r_{obs}<r_{bar}$). Figure \ref{fig:1} top left panel shows data in $g2$-space, normalized according to equations \eqref{norm2} and \eqref{norm22}, for the $11$ gas-dominated galaxies.
Figure \ref{fig:1} middle left panel and bottom left panel show the top left panel split up into leftward (middle left panel) and rightward (bottom left panel) galaxies. In order to quantify the overall geometry of data in normalized $g2$-space, data is binned into $8$ bins at large radii and $8$ at small radii, each defined by an interval of $\hat{g}_{bar}$. The $7$ bins at small and large radii with lowest values in $\hat{g}_{bar}$ have a width of $0.125$. The two bins at largest $\hat{g}_{bar}$ contain points for which $\hat{g}_{bar}>0.875$. The average within the $m$'th bin is given by
\begin{equation}
(\braket{\hat{g}_{bar}}_{m},\braket{\hat{g}_{obs}}_{m}),
\end{equation}
where
\begin{eqnarray}
\braket{\hat{g}_{bar}}_{m}=\frac{1}{N_{m}}\sum_{n=1}^{N_{gal}}\sum_{j\in m_n}\hat{g}_{bar}(r_{j,n},r_{bar\pm 1,n}),\\
\braket{\hat{g}_{obs}}_{m}=\frac{1}{N_{m}}\sum_{n=1}^{N_{gal}}\sum_{j\in m_n}\hat{g}_{obs}(r_{j,n},r_{bar\pm 1,n}),\\
\label{bin2}\nonumber
\end{eqnarray}
where $m_n$ denote the data points of the $n$'th galaxy that falls within the $m$'th bin and $N_m$ denote all points within the $m$'th bin. The derivation of the uncertainty of the binned plots is reviewed in appendix \ref{sec:details}. The right column in figure \ref{fig:1} show binned plots of the left column. From the top right panel it is clear that the overall normalized $g2$-space geometry of gas-dominated galaxies is rightward, as opposed to the overall $\sim\,$nowhere geometry found in \citep{Frandsen:2018ftj} (shown in figure \ref{fig:5} left panel). The overall geometry in normalized $g2$-space progressively approach a nowhere geometry, as found in \citep{Frandsen:2018ftj}, as the constraint on the galaxies being gas-dominated is relaxed from $80\%$ to $50\%$. Hence as the dependence on the mass-to-light ratios is increased (the gas fraction is decreased), the overall geometry is taken from a nowhere to a rightward geometry. Assuming there is nothing dynamically unique (in relation to the $g2$-space geometry) about gas dominated galaxies\footnote{I thank Federico Lelli for pointing out that the rotational velocity is difficult to measure accurately when the circular and non-circular speeds are of the same order -- which often is the case for data points at the innermost radii. In this regard, I would like to stress that the results of this article are off course given the integrity of the data used. That being said, the data points discussed here are the same ones discussed in the cusp/core debate -- with leftward/rightward galaxies representing cusp/core galaxies, respectively -- and so data are not stretched in an unconventional manner. The issue is discussed further in appendix \ref{app:cut} with the consensus being that it does not play a significant role.}, this can be interpreted as indicating that the overall nowhere geometry found in \citep{Frandsen:2018ftj} is an artefact of a significant radial dependence of the mass to light ratios (not considered in \citep{Frandsen:2018ftj}, but discussed as a caveat) and by extension that the true overall geometry is in fact rightward. From this information it can be inferred that the underlying radial dependence should be such that an overall nowhere geometry is taken to an overall rightward, meaning that the radial dependence of the mass to light ratios should be increasing significantly toward small radii. This is what is expected from a radial dependence arising from metallicity gradients, e.g. as discussed in \citep{Portinari:2009ap}. Inspired by \citep{Portinari:2009ap} 
\begin{equation}
\tilde{\Upsilon}^{d}_i(r)=0.5e^{-(\frac{5}{4}(0.3)^3+0.13)(\frac{r}{r^d_i}-1)}
\label{Yd}
\end{equation}
can be adopted instead of a constant disk mass to light ratio. Adopting equation \eqref{Yd} to some degree shift the overall geometry of \emph{all} $152$ galaxies from nowhere to rightward (see figure \ref{fig:5}). Since equation \eqref{Yd} is just one suggestion of a radial dependence of only the disk mass to light ratio, it is included just to show that the conclusions drawn from the gas-dominated galaxies are consistent with some considerations in the discussion surrounding a potential radial dependence of the mass to light ratios and that the required correction inferred from the gas dominated galaxies is consistent in terms of magnitude.\newline
Information about the overall geometry of data in normalized $g2$-space is highly significant and can be used to reject and shape solutions to the missing mass problem. For example, MOND modified gravity in the Brada-Milgrom formulation can (to first order \citep{Frandsen:2018ftj}) only accommodate a rightward geometry, meaning that according to this interpretation the Brada-Milgrom approximation of MOND modified gravity cannot be ruled out by referring to the average geometry. Another example is that of NFW dark matter which unequivocally give rise to a leftward $g2$-space geometry \citep{Frandsen:2018ftj}. Hence, determining whether a subset of galaxies display a leftward geometry or not can be used to limit the potential use of the NFW profile - and by extension related physics.\newline
The determination of the significance of the different sub-categories (rightward, leftward and nowhere) is also highly significant in relation to the cusp/core debate for dark matter haloes since leftward galaxies represent a cusped profile whereas rightward galaxies represent a cored profile.
As a last comment it should be noted that, as discussed in \citep{Frandsen:2018ftj}, incorporating a radial dependence into the mass to light ratios does not change the conclusions drawn in \cite{Frandsen:2018ftj} which pertains to rejecting MOND modified inertia based on investigating whether $r_{bar}=r_{obs}$ - as MOND modified inertia in general predicts \citep{Frandsen:2018ftj,Petersen:2017klw}. In the notation of this article $r_{obs}=r_{bar}$ can be tested (\emph{for all galaxies}) similarly to what is done in \citep{Frandsen:2018ftj} by considering the average of equations \eqref{norm22} and \eqref{norm2}, symmetric with respect to the points included in the sum over $s$
\begin{equation}
\begin{split}
\braket{\hat{g}_{obs}(r_{obs \pm 1,i},r_{bar\pm 1,i})}&\equiv \frac{1}{N_i}\sum_{k=obs\pm 1}\hat{g}_{obs}(r_{k,i},r_{bar\pm 1,i}),\\
\end{split}
\label{1}
\end{equation}
\begin{equation}
\begin{split}
\braket{\hat{g}_{bar}(r_{obs \pm 1,i},r_{bar\pm 1,i})}&\equiv \frac{1}{N_i}\sum_{k=obs\pm 1}\hat{g}_{bar}(r_{k,i},r_{bar\pm 1,i}).\\
\end{split}
\label{2}
\end{equation}
Using $\tilde{\Upsilon}_i^b=0.7$, $\tilde{\Upsilon}^{d}_i$ given by equation \eqref{Yd} and propagating the uncertainty similarly to what is shown in appendix \ref{sec:details}
\begin{align}
&\braket{\hat{g}_{obs}(r_{obs \pm 1,i},r_{bar\pm 1,i})}=1.302\pm 0.039,\nonumber\\
&\braket{\hat{g}_{bar}(r_{obs \pm 1,i},r_{bar\pm 1,i})}=0.843\pm 0.003.\nonumber\\
\label{3}
\end{align}
Both quantities in equation \eqref{3} yield more than $8\sigma$ deviation from the prediction of MOND modified inertia (unity). 

\section{Summary and conclusions}
It this study it has been shown that the geometry of gas-dominated galaxies (as defined in equation \eqref{gdom}) in $g2$-space is consistent with an overall rightward geometry ($r_{obs}>r_{bar}$ characteristic of e.g. pseudo-isothermal dark matter and MOND modified gravity). This is in contrast to the overall geometry of all galaxies found in \citep{Frandsen:2018ftj} (and reproduced in the left panel of figure \ref{fig:5}) which is largely consistent with a nowhere geometry (using $\tilde{\Upsilon}^{d}_i=0.5$ and $\tilde{\Upsilon}^{b}_i=0.7$). To be clear; this difference is caused by the galaxies considered in the two analysis'. The overall geometry in normalized $g2$-space progressively approach a nowhere geometry, as found in \citep{Frandsen:2018ftj}, as the constraint on the galaxies being gas-dominated is relaxed. Hence as the dependence on the mass-to-light ratio(s) is increased (the gas fraction - as defined in equation \eqref{gdom} - is decreased), the overall geometry is taken from a nowhere to a rightward geometry. Assuming there is nothing dynamically unique\footnote{See footnote 3 and appendix \ref{app:cut} for a discussion of this point.} (in relation to the $g2$-space geometry) about gas dominated galaxies, this can be interpreted as indicating that the overall nowhere geometry found in \citep{Frandsen:2018ftj} is an artefact of a significant radial dependence of the mass to light ratios and by extension that the true overall geometry is in fact rightward. The inferred radial dependence of the mass to light ratios should be such that they significantly increase toward smaller radii. This is consistent with what is expected from a radial dependence of the mass to light ratios coming from metallicity gradients \citep{Portinari:2009ap}. A particular radial dependency of the disk mass to light ratio - suggested by \citep{Portinari:2009ap} - has been investigated (equation \eqref{Yd}) and it is found that applying this particular profile results in a rightward shift in the overall geometry of the required form and magnitude (see figure \ref{fig:5} and compare to figure \ref{fig:1} top right panel). Hence, the radial dependency of the mass to light ratios inferred from comparing the $g2$-space geometry of the gas dominated galaxies to that of all galaxies is consistent - both in terms of tendency and magnitude - with (at least) some considerations in the discussion surrounding a potential radial dependency of the mass to light ratios.\newline
The potential absence of leftward and nowhere galaxies would be an important information for both dark matter and modified gravity model builders. If there is a radial dependency of the mass to light ratios as suggested the amount of leftward (and nowhere of course) galaxies would (at least) be brought significantly down (from the already relatively low $33/152$ galaxies). As leftward galaxies represent cusped dark matter profiles, this would mean that the applicability of the cuspy profiles would be limited. This information can by extension be used to infer properties on the required particle dark matter and shape the model building process of both dark matter and modified gravity (as modified gravity models consistent with cuspy dark matter on galactic scales would be disfavored).\newline 
Lastly, it has been shown that the considered radial dependency of the disk mass to light ratio does not change the conclusions drawn in \cite{Frandsen:2018ftj} which pertains to rejecting MOND modified inertia.\bigskip

{\bf Acknowledgments:}
I thank Mads Frandsen for comments on the draft as well as bringing the work of \cite{Salucci:2016vxb} to my attention. I thank Federico Lelli for comments on the draft and discussions of difficulties in measuring data points at small radii.\newline 
Lastly I acknowledge the partial funding from The Council For Independent Research, grant number DFF 6108-00623. The CP3-Origins center is partially funded by the Danish National Research Foundation, grant number DNRF90.\bigskip

\begin{appendices}

\appendixpage
\noappendicestocpagenum
\addappheadtotoc

\section{Rotation Curve Plots}
\label{app:plots}
\begin{figure}[H]
	\centering
	\captionsetup{width=0.5\textwidth}
	\begin{subfigure}{0.3\textwidth}
		\includegraphics[width=1\textwidth]{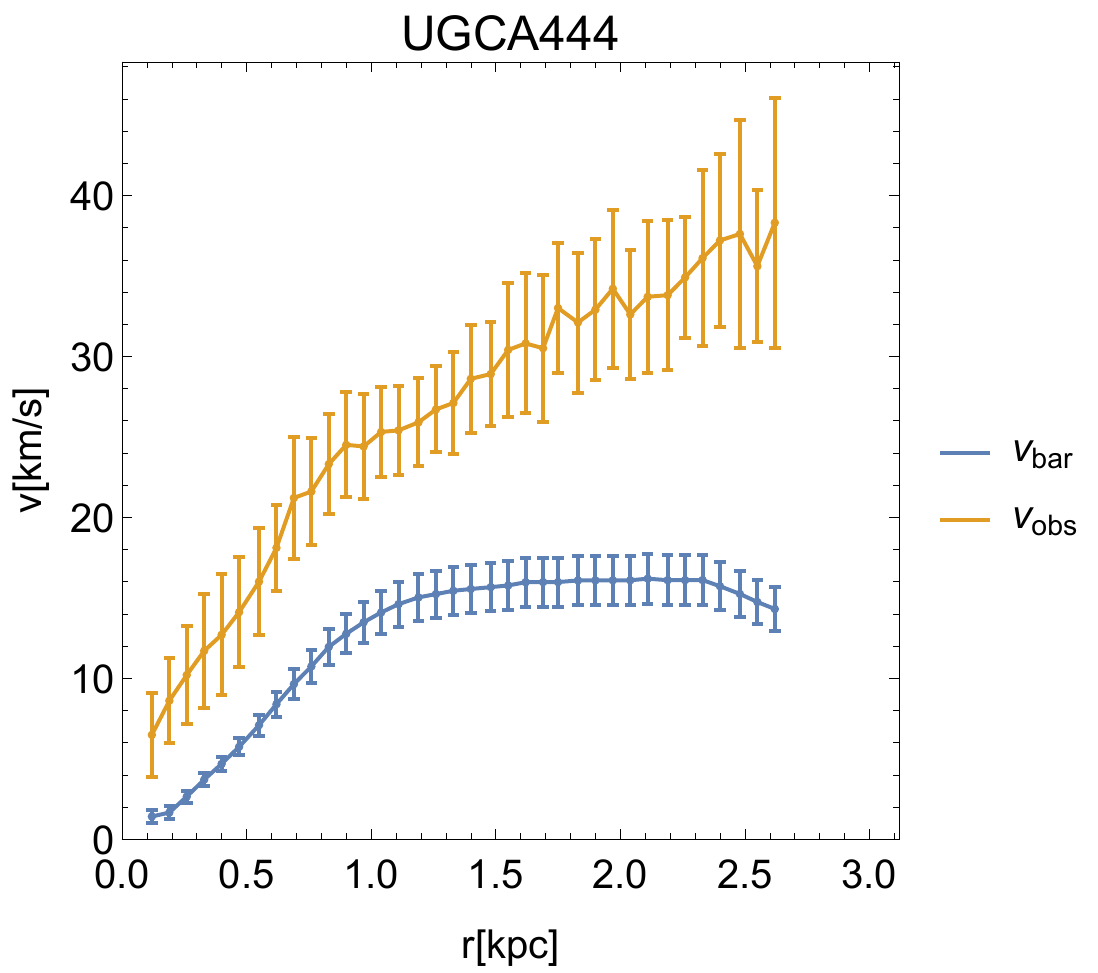}
	\end{subfigure}
	\begin{subfigure}{0.3\textwidth}
		\includegraphics[width=1\textwidth]{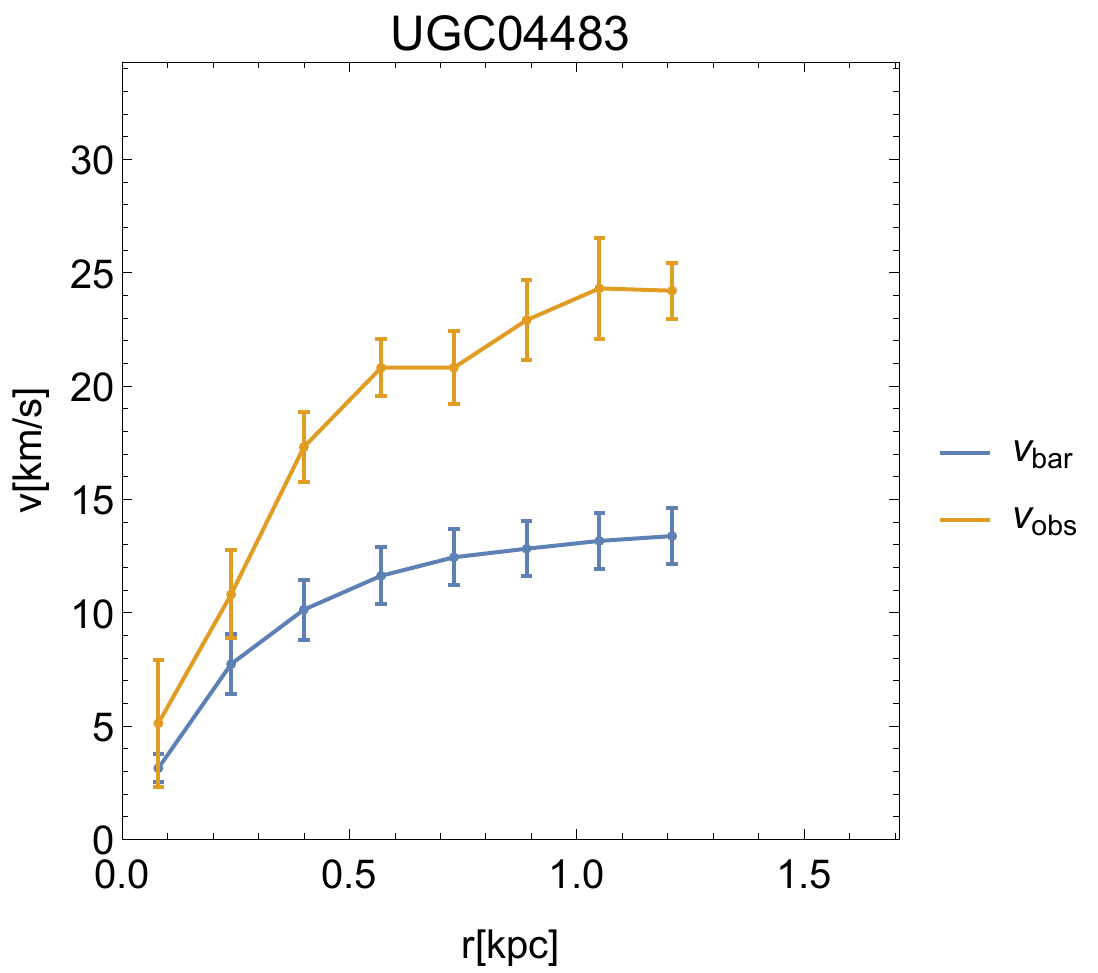}
	\end{subfigure}
	\begin{subfigure}{0.3\textwidth}
		\includegraphics[width=1\textwidth]{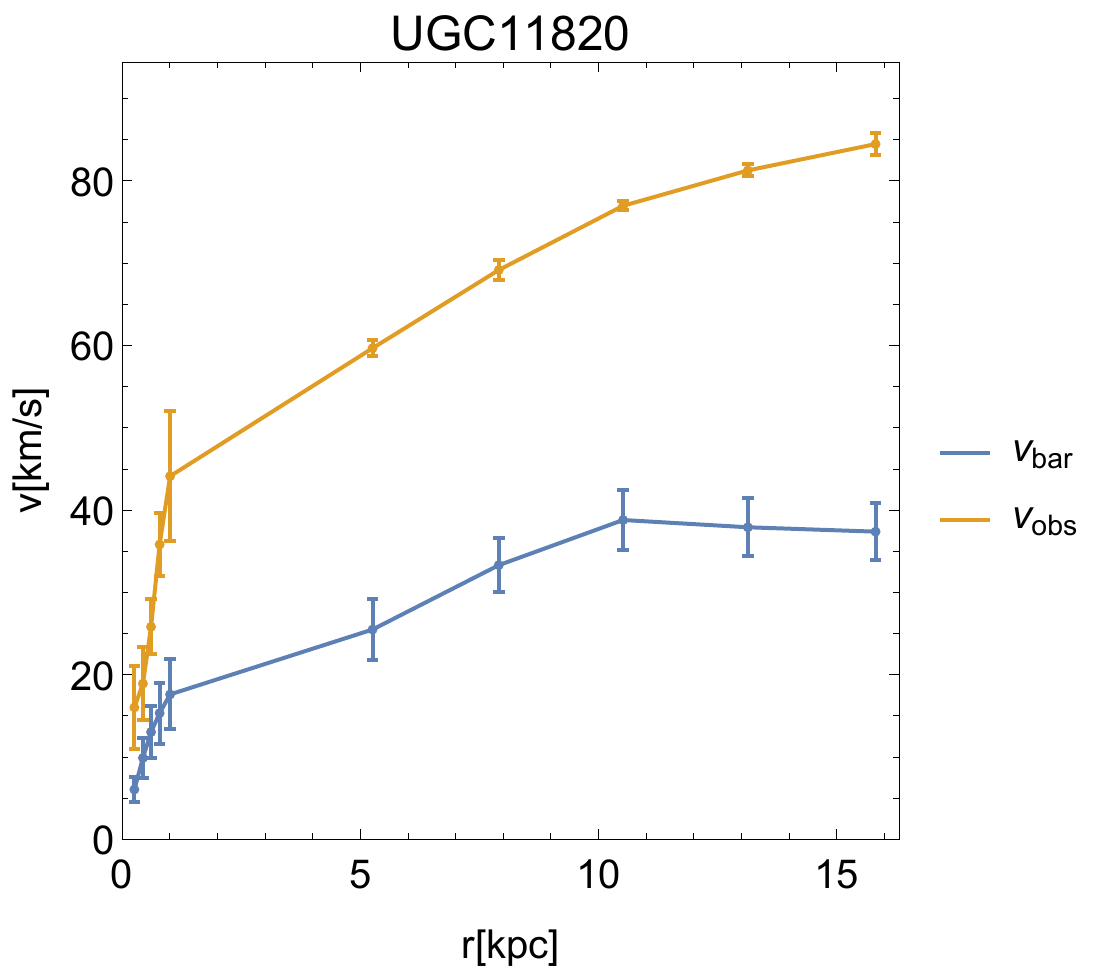}
	\end{subfigure}
\end{figure}

\begin{figure}[H]
	\centering
	\captionsetup{width=0.5\textwidth}
	\begin{subfigure}{0.3\textwidth}
		\includegraphics[width=1\textwidth]{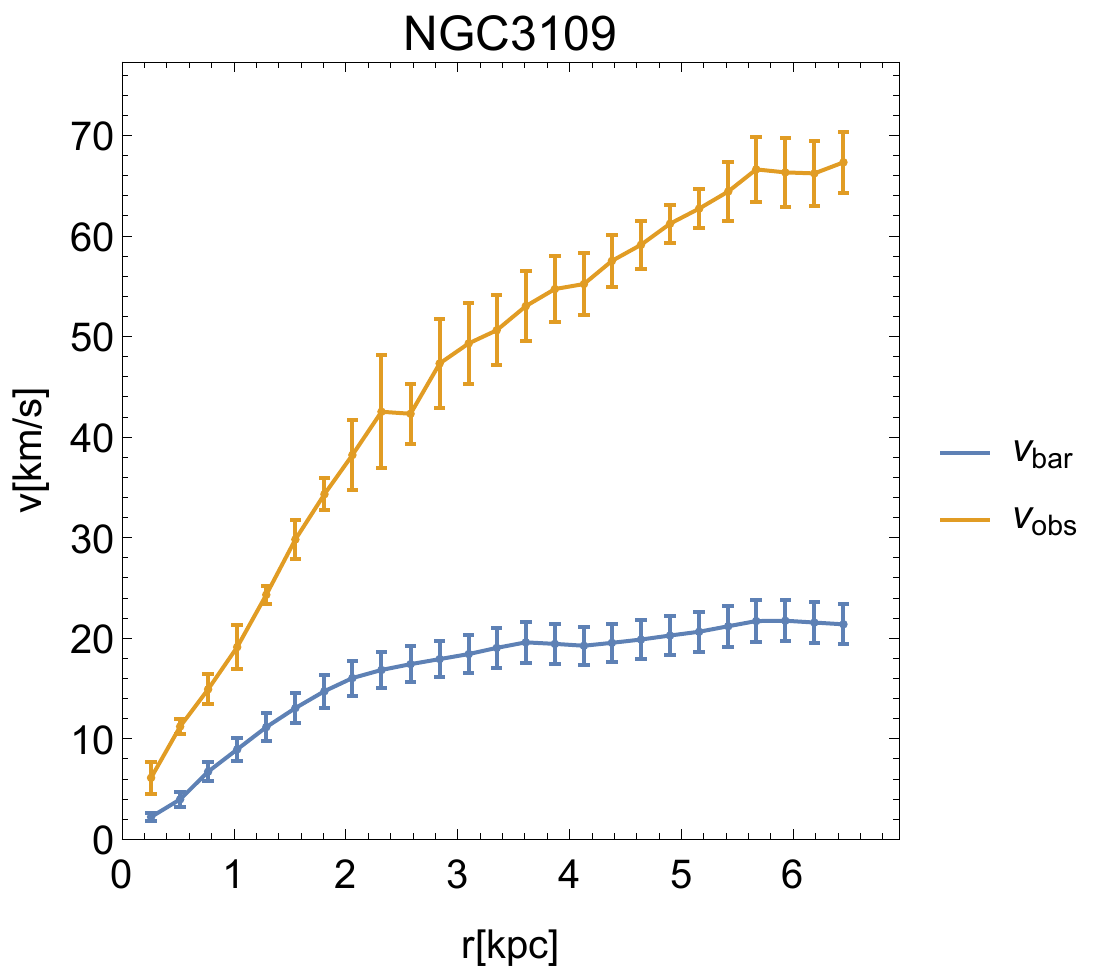}
	\end{subfigure}
	\begin{subfigure}{0.3\textwidth}
		\includegraphics[width=1\textwidth]{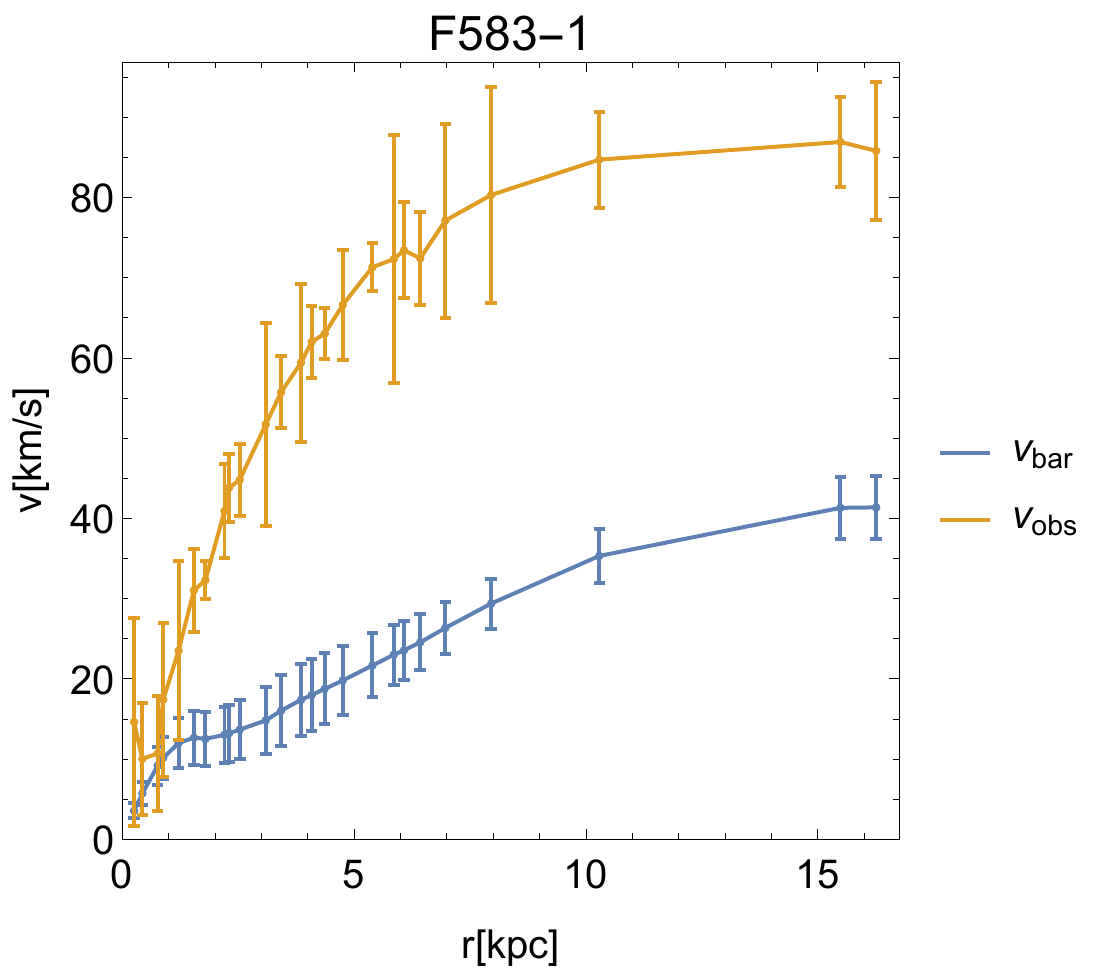}
	\end{subfigure}
	\begin{subfigure}{0.3\textwidth}
		\includegraphics[width=1\textwidth]{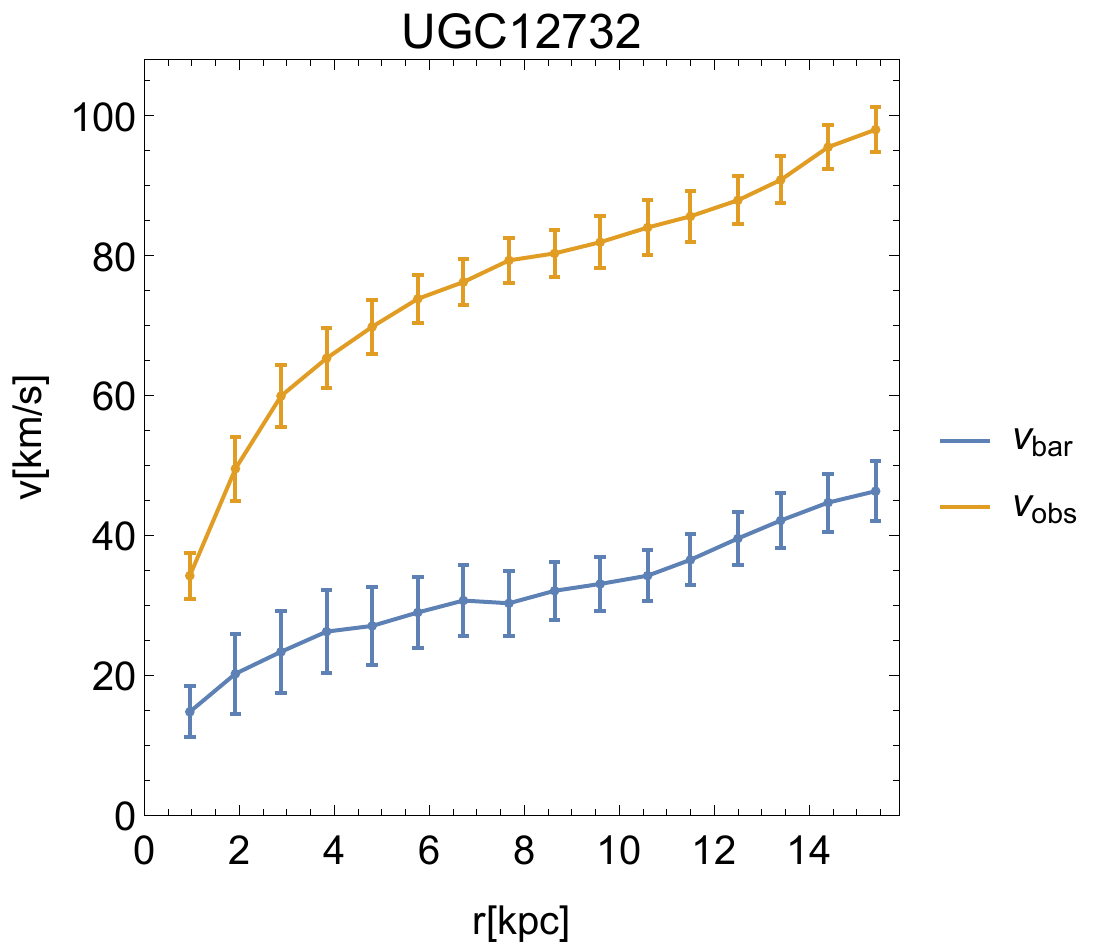}
	\end{subfigure}
	\begin{subfigure}{0.3\textwidth}
		\includegraphics[width=1\textwidth]{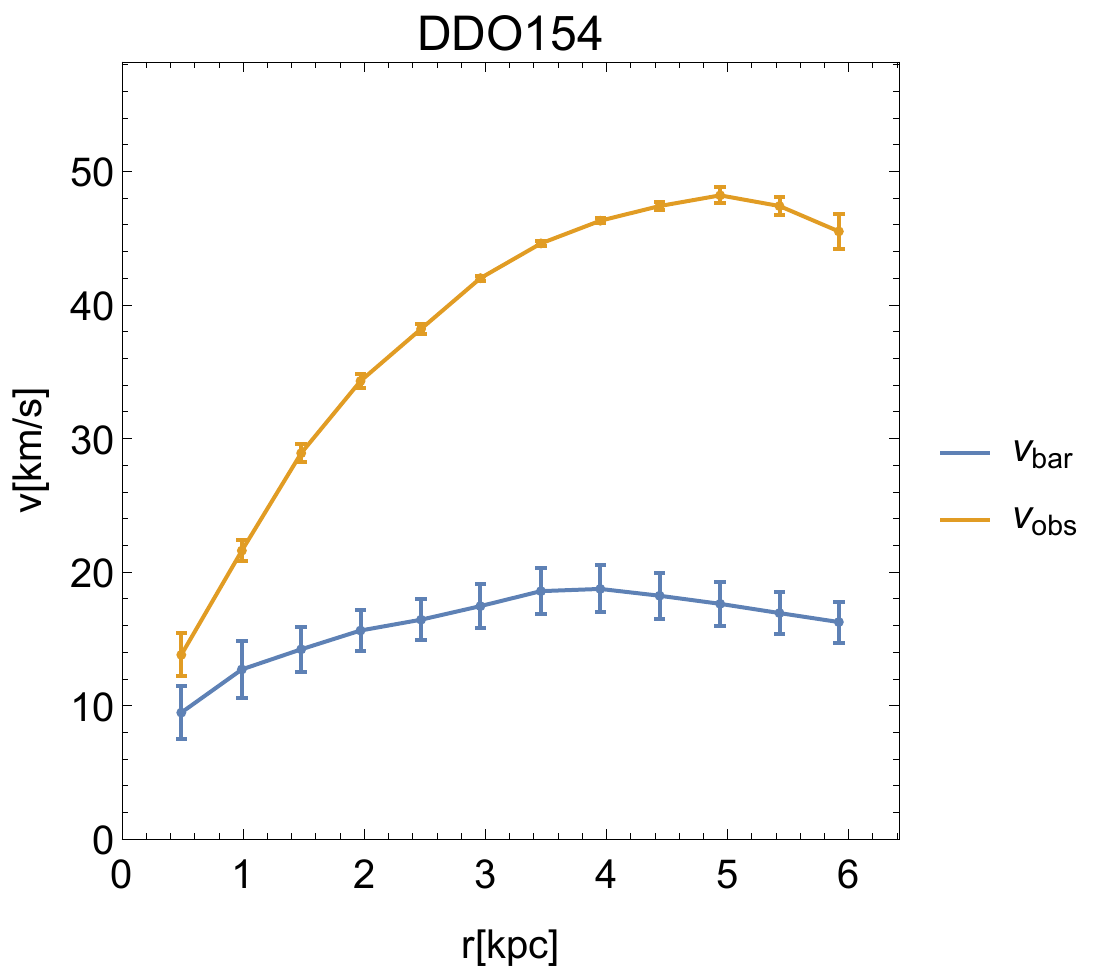}
	\end{subfigure}
	\begin{subfigure}{0.3\textwidth}
		\includegraphics[width=1\textwidth]{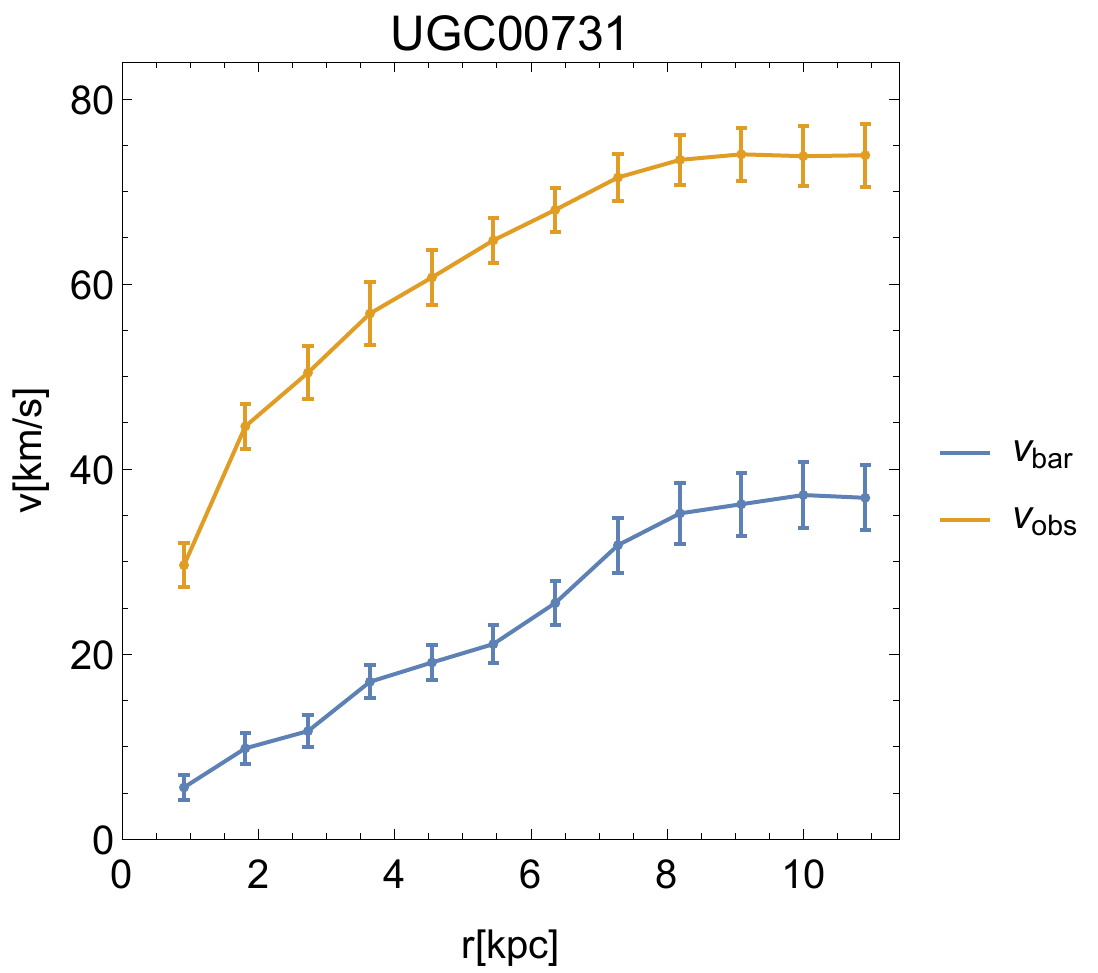}
	\end{subfigure}
\end{figure}

\begin{figure}[H]
	\centering
	\captionsetup{width=0.5\textwidth}
	\begin{subfigure}{0.3\textwidth}
		\includegraphics[width=1\textwidth]{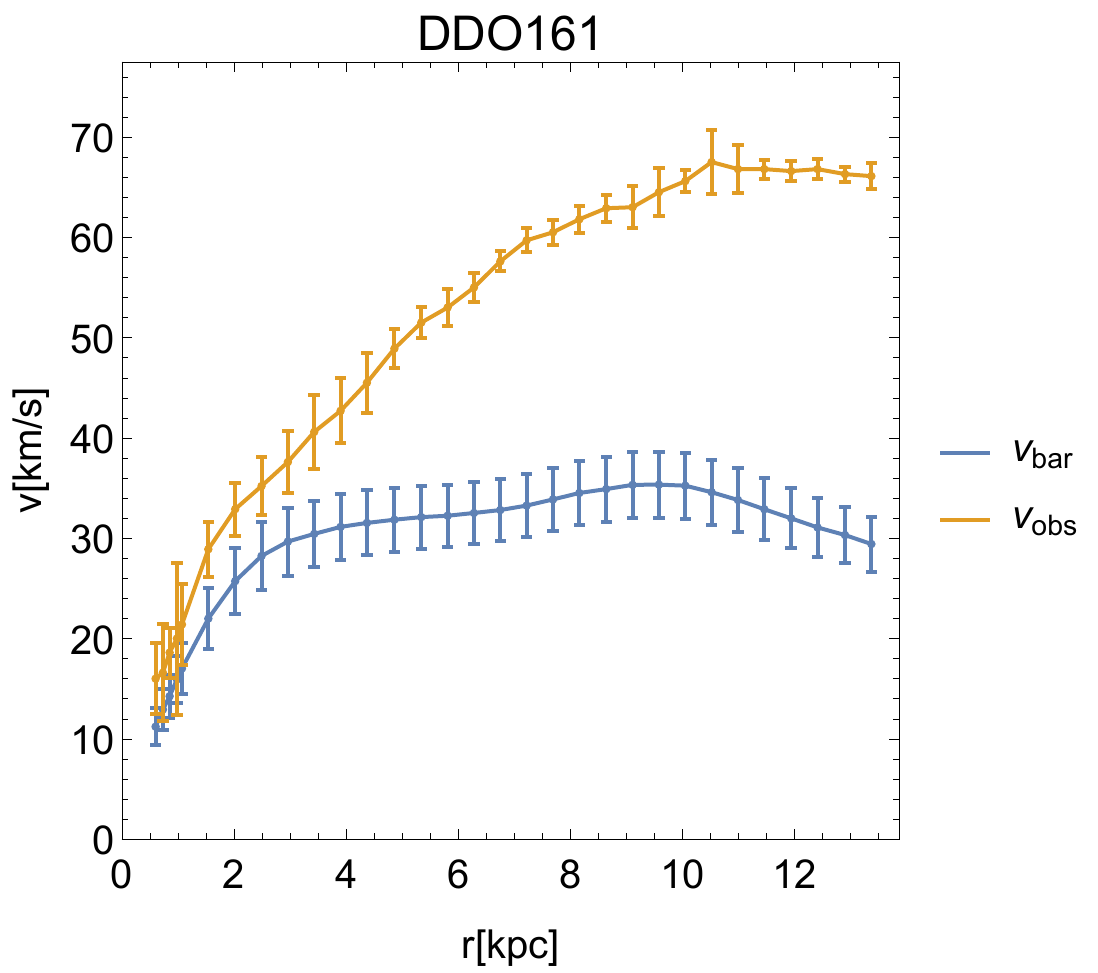}
	\end{subfigure}
	\begin{subfigure}{0.3\textwidth}
		\includegraphics[width=1\textwidth]{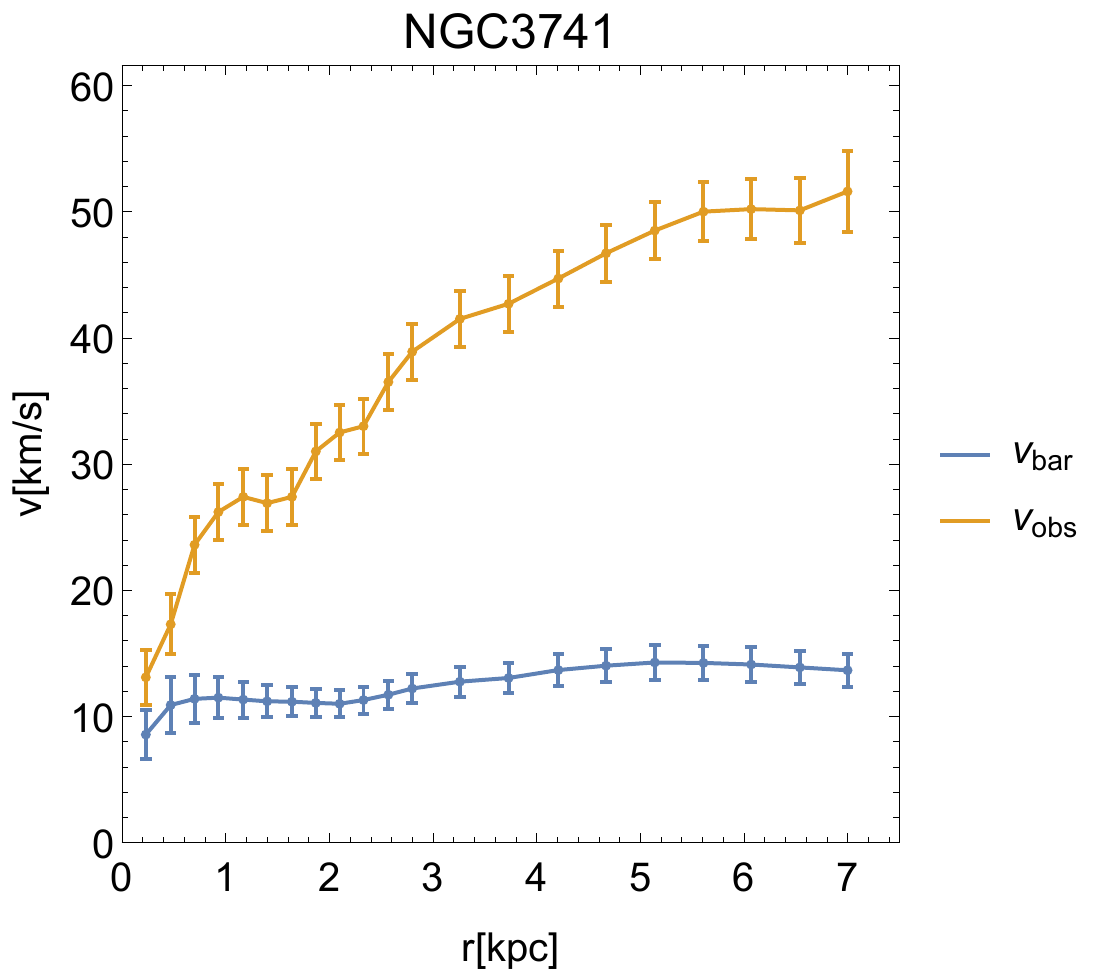}
	\end{subfigure}
\end{figure}

\begin{figure}[H]
	\centering
	\captionsetup{width=0.5\textwidth}
	\begin{subfigure}{0.3\textwidth}
		\includegraphics[width=1\textwidth]{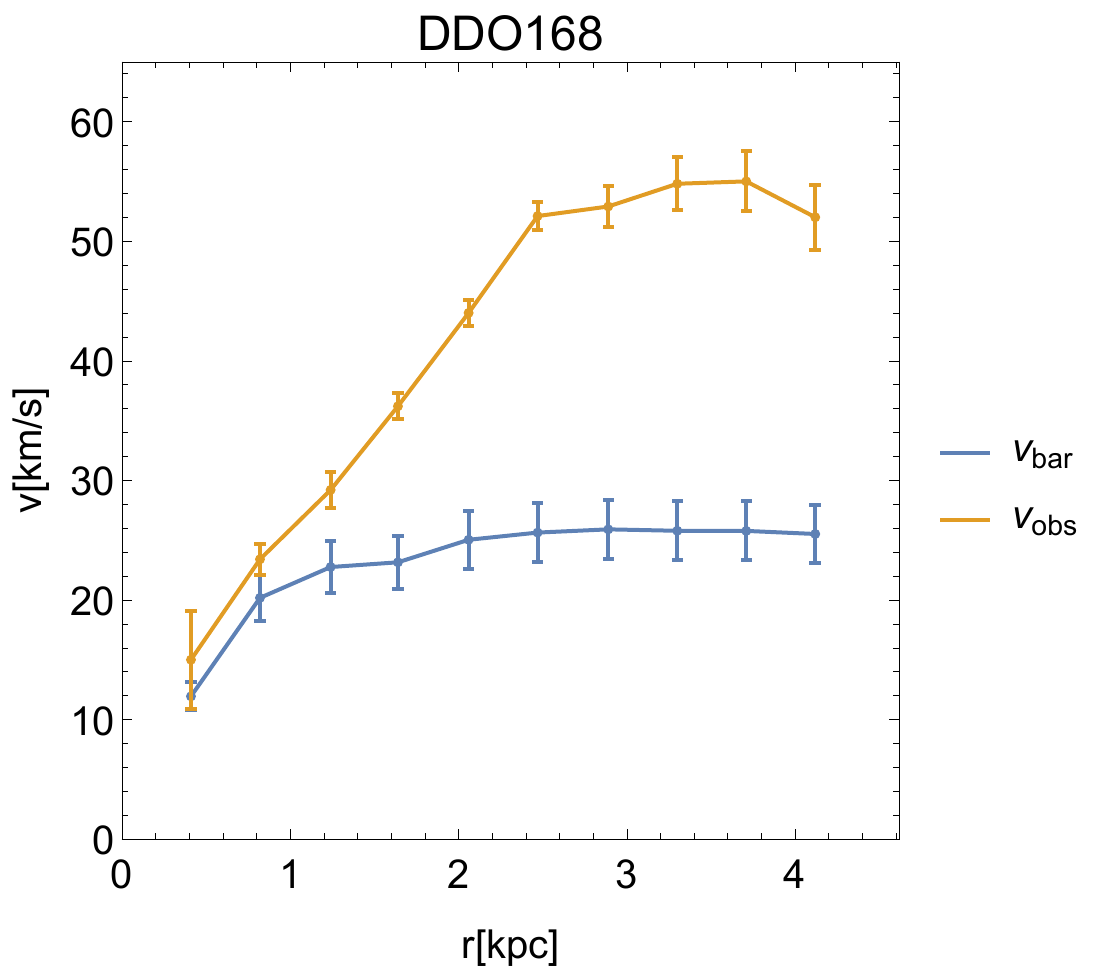}
	\end{subfigure}
\end{figure}

\section{Binned $g2$-space Plots}
\label{sec:details}
The average within the $m$'th bin is given by
\begin{equation}
(\braket{\hat{g}_{bar}}_{m},\braket{\hat{g}_{obs}}_{m}),
\end{equation}
where
\begin{eqnarray}
\braket{\hat{g}_{obs}}_{m}=\frac{1}{N_{m}}\sum_{n=1}^{N_{gal}}\sum_{j\in m_n}\hat{g}_{obs}(r_{j,n},r_{bar\pm 1,n}),\\
\braket{\hat{g}_{bar}}_{m}=\frac{1}{N_{m}}\sum_{n=1}^{N_{gal}}\sum_{j\in m_n}\hat{g}_{bar}(r_{j,n},r_{bar\pm 1,n}),
\end{eqnarray}
where $m_n$ denotes the data points of the $n$'th galaxy that falls within the $m$'th bin and $N_m$ denotes all points within the $m$'th bin. The uncertainty of $\braket{\hat{g}_{obs}}_{m}$ is found by propagating the uncertainty of $v_{obs}$
\begin{equation}
\delta \braket{\hat{g}_{obs}}_{m}=\sqrt{\sum_{i=1}^{N_{gal}}\sum_{k\in gal}\bigg(\frac{\partial \braket{\hat{g}_{obs}}_{m}}{\partial v_{obs}(r_{k,i})}\delta v_{obs}(r_{k,i})\bigg)^2}.
\label{pro7}
\end{equation}
Using
\begin{widetext}
	\begin{equation}
	\frac{\partial \braket{\hat{g}_{obs}}_{m}}{\partial v_{obs}(r_{k,i})}=\frac{1}{N_m}\sum_{j\in m_i}\frac{2g_{obs}(r_{j,i})}{\braket{g_{obs}(r_{bar,i})}}\bigg[\frac{\delta_{k,j}}{v_{obs}(r_{j,i})}-\frac{\sum_{s=bar_i\pm 1}\frac{v_{obs}(r_{s,i})}{r_{s,i}}\delta_{s,k}}{N_i\braket{g_{obs}(r_{bar,i})}}
	\bigg]
	\end{equation}
	in equation \eqref{pro7} $(\delta \braket{\hat{g}_{obs}}_{m})^2$ is determined to be
	\begin{equation}
	\begin{split}
	(\delta \braket{\hat{g}_{obs}}_{m})^2=&\frac{1}{N_m^2}\sum_{i=1}^{N_{gal}}\bigg[\sum_{k\in m_i}\bigg(\frac{2g_{obs}(r_{k,i})}{\braket{g_{obs}(r_{bar,i})}}\frac{\delta v_{obs}(r_{k,i})}{v_{obs}(r_{k,i})}\bigg)^2\\
	&+\sum_{s=bar_i\pm 1}\bigg(\frac{2g_{obs}(r_{s,i})}{\braket{g_{obs}(r_{bar,i})}}\frac{\delta v_{obs}(r_{s,i})}{v_{obs}(r_{s,i})}\frac{1}{N_i}\sum_{j\in m_i}\frac{g_{obs}(r_{j,i})}{\braket{g_{obs}(r_{bar,i})}}\bigg)^2\\
	&-2\sum_{u=k\cap s}\bigg(\frac{2g_{obs}(r_{u,i})}{\braket{g_{obs}(r_{bar,i})}}\frac{\delta v_{obs}(r_{u,i})}{v_{obs}(r_{u,i})}\bigg)^2\frac{1}{N_i}\sum_{j\in m_i}\frac{g_{obs}(r_{j,i})}{\braket{g_{obs}(r_{bar,i})}}\bigg],\\
	\end{split}
	\label{pro8}
	\end{equation}
	where $u$ run over the intersection between $k\in m_i$ and $s=bar_i\pm 1$. For $\braket{\hat{g}_{bar}}_{m}$
	\begin{equation}
	\delta \braket{\hat{g}_{bar}}_{m}=\sqrt{\sum_{i=1}^{N_{gal}}\bigg[\bigg(\frac{\partial \braket{\hat{g}_{bar}}_{m}}{\partial m^g_i}\delta m_i^g\bigg)^2+\sum_{w=d,b}\bigg(\frac{\partial \braket{\hat{g}_{bar}}_{m}}{\partial \tilde{\Upsilon}_i^w}\delta \tilde{\Upsilon}_i^w\bigg)^2\bigg]}.
	\label{pro9}
	\end{equation}
	Taking $\frac{|v_g(r_{j,i})|v_g(r_{j,i})}{r_{j,i}}\propto m_i^g$ reveals
	\begin{equation}
	\begin{split}
	&\frac{\partial \braket{\hat{g}_{bar}}_{m}}{\partial m_i^g}=\frac{1}{N_mm_i^g}\sum_{j\in m_i}\frac{g_{bar}(r_{j,i})}{\braket{g_{bar}(r_{bar,i})}}\bigg[\frac{|v_g(r_{j,i})|v_g(r_{j,i})}{v_{bar}^2(r_{j,i})}-\frac{\sum_{s=bar_i\pm 1}\frac{|v_g(r_{s,i})|v_g(r_{s,i})}{r_{s,i}}}{N_i\braket{g_{bar}(r_{bar,i})}}\bigg],\\
	&\frac{\partial \braket{\hat{g}_{bar}}_{m}}{\partial \tilde{\Upsilon}_i^w}=\frac{1}{N_m}\sum_{j\in m_i}\frac{g_{bar}(r_{j,i})}{\braket{g_{bar}(r_{bar,i})}}\bigg[\frac{v_w^2(r_{j,i})}{v_{bar}^2(r_{j,i})}-\frac{\sum_{s=bar_i\pm 1}\frac{v_w^2(r_{s,i})}{r_{s,i}}}{N_i\braket{g_{bar}(r_{bar,i})}}\bigg].\\
	\end{split}
	\label{pro11}
	\end{equation}
	Using equation \eqref{pro11} in equation \eqref{pro9}, $(\delta \braket{\hat{g}_{bar}}_{m})^2$ is determined to be
	\begin{equation}
	\begin{split}
	(\delta \braket{\hat{g}_{bar}}_{m})^2=\frac{1}{N_m^2}\sum_{i\in gal}\bigg[&\bigg(\frac{\delta m_i^g}{m_i^g}\sum_{j\in m_i}\frac{g_{bar}(r_{j,i})}{\braket{g_{bar}(r_{bar,i})}}\bigg[\frac{|v_g(r_{j,i})|v_g(r_{j,i})}{v_{bar}^2(r_{j,i})}-\frac{\sum_{s=bar_i\pm 1}\frac{|v_g(r_{s,i})|v_g(r_{s,i})}{r_{s,i}}}{N_i\braket{g_{bar}(r_{bar,i})}}\bigg]\bigg)^2\\
	&+
	\sum_{w=d,b}\bigg(\delta \tilde{\Upsilon}_i^w\sum_{j\in m_i}\frac{g_{bar}(r_{j,i})}{\braket{g_{bar}(r_{bar,i})}}\bigg[\frac{v_w^2(r_{j,i})}{v_{bar}^2(r_{j,i})}-\frac{\sum_{s=bar_i\pm 1}\frac{v_w^2(r_{s,i})}{r_{s,i}}}{N_i\braket{g_{bar}(r_{bar,i})}}\bigg]\bigg)^2\bigg],
	\end{split}
	\label{pro10}
	\end{equation}
	where $\delta m_i^g=0.1m_i^g$ and $\delta \tilde{\Upsilon}_i^w=0.25\tilde{\Upsilon}_i^w$.\newline\newline\newline\newline\newline
	
\end{widetext}

\section{Discussion Related to footnote 3}
\label{app:cut}
\begin{figure*}[!htb]
	\centering
	\captionsetup{width=1\textwidth}
	\begin{subfigure}{0.3\textwidth}
		\includegraphics[width=1\textwidth]{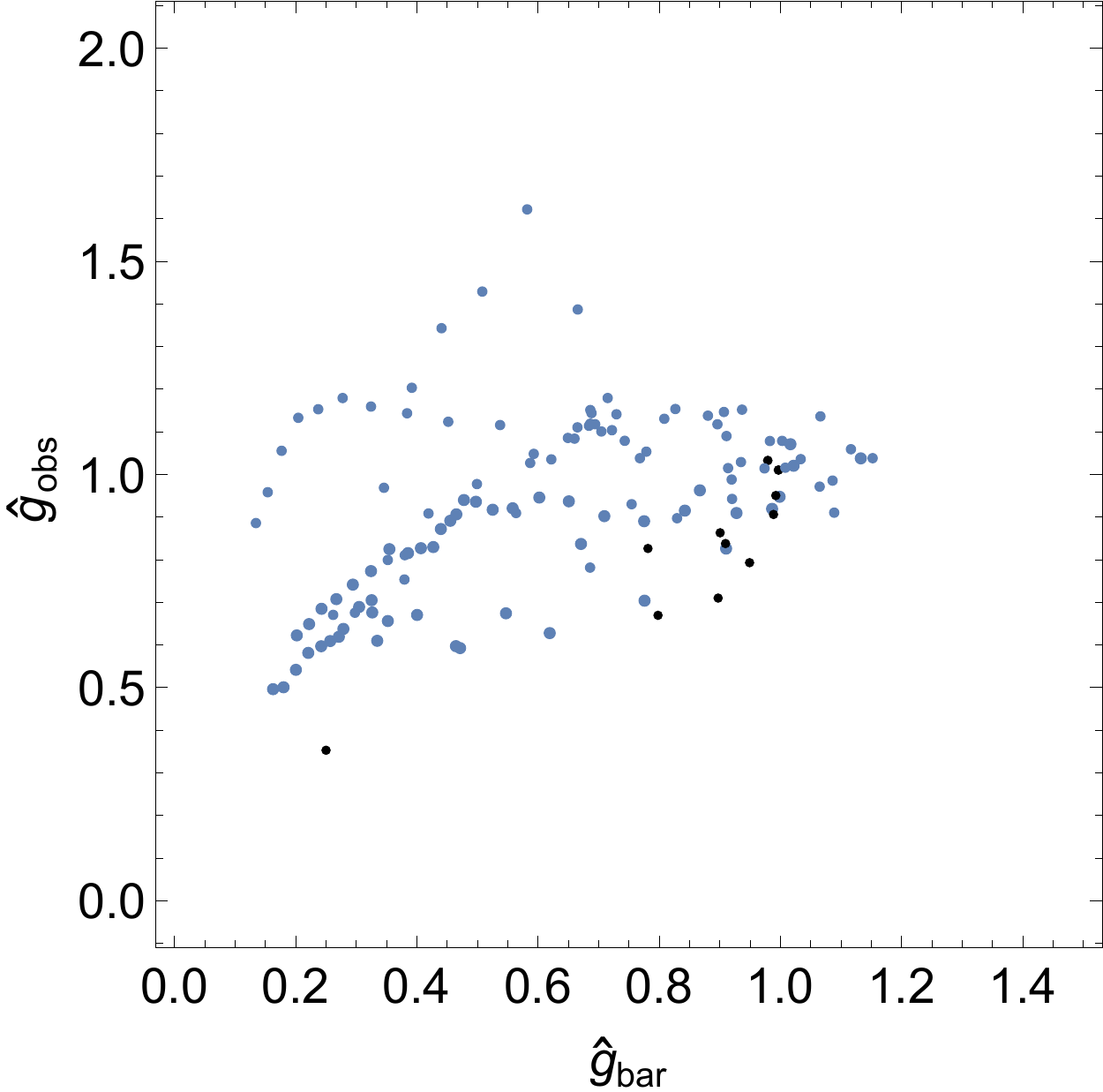}
	\end{subfigure}
	\begin{subfigure}{0.3\textwidth}
		\includegraphics[width=1\textwidth]{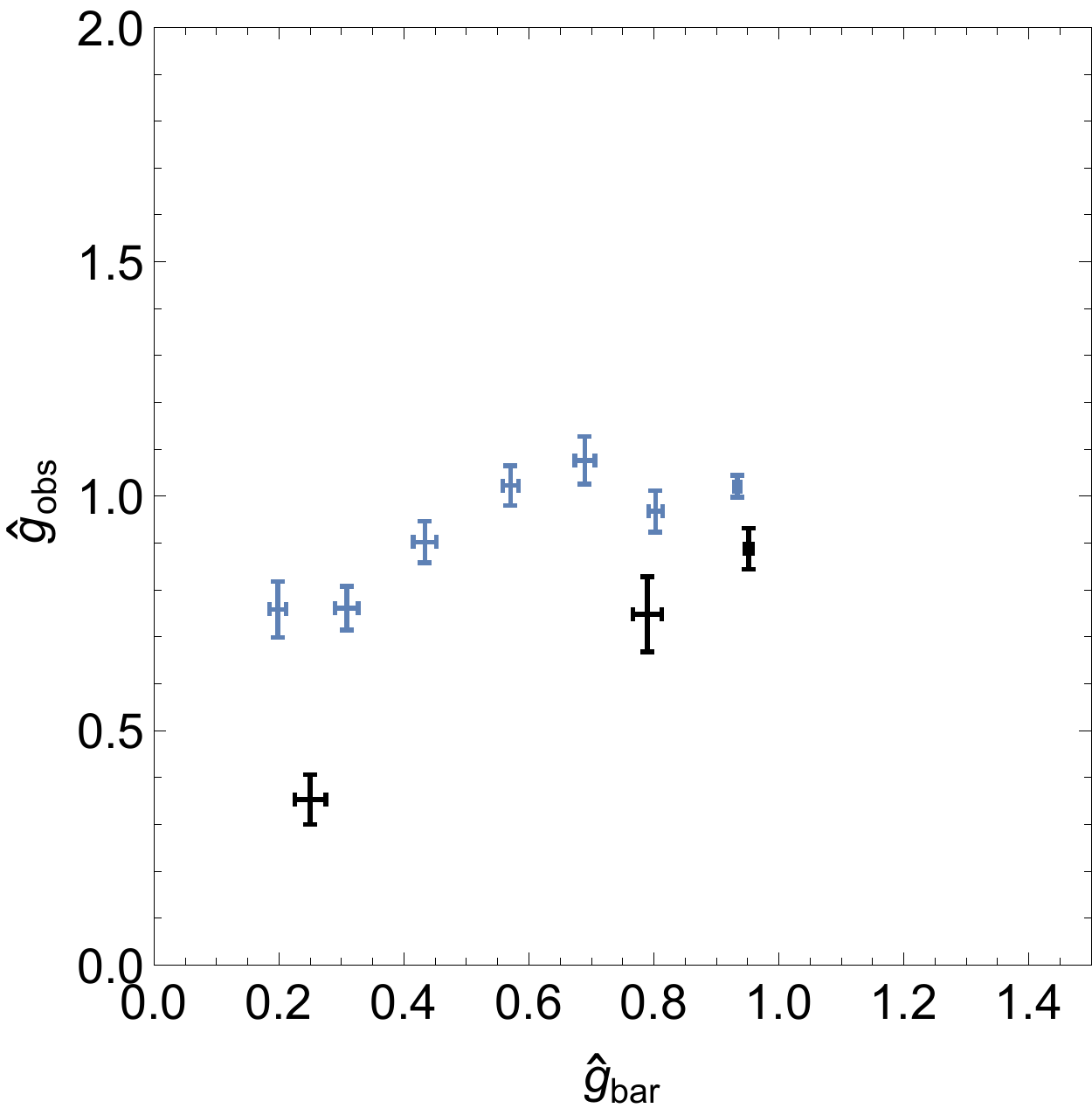}
	\end{subfigure}\\
	\begin{subfigure}{0.3\textwidth}
		\includegraphics[width=1\textwidth]{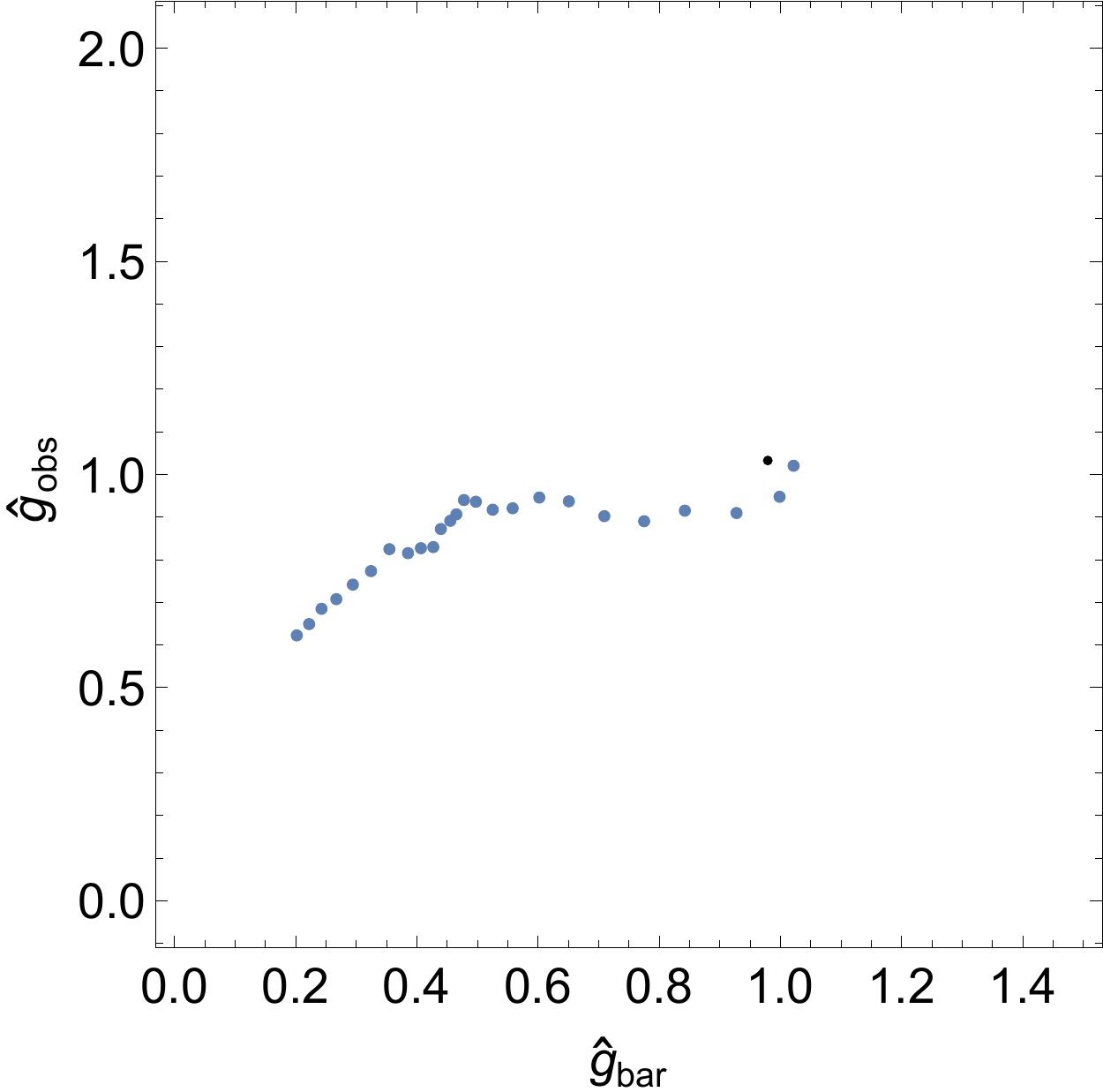}
	\end{subfigure}
	\begin{subfigure}{0.3\textwidth}
		\includegraphics[width=1\textwidth]{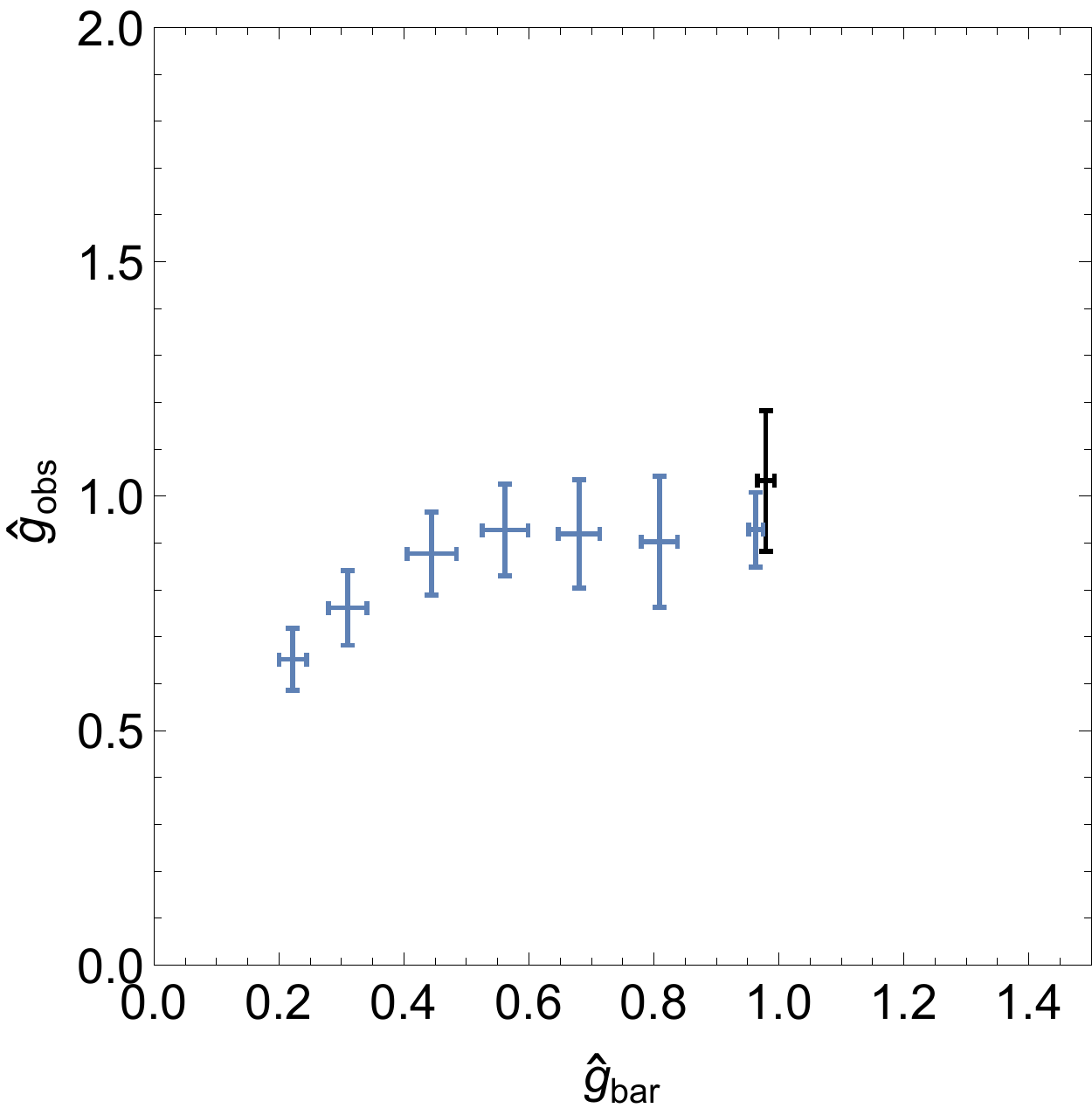}
	\end{subfigure}\\
	\begin{subfigure}{0.3\textwidth}
		\includegraphics[width=1\textwidth]{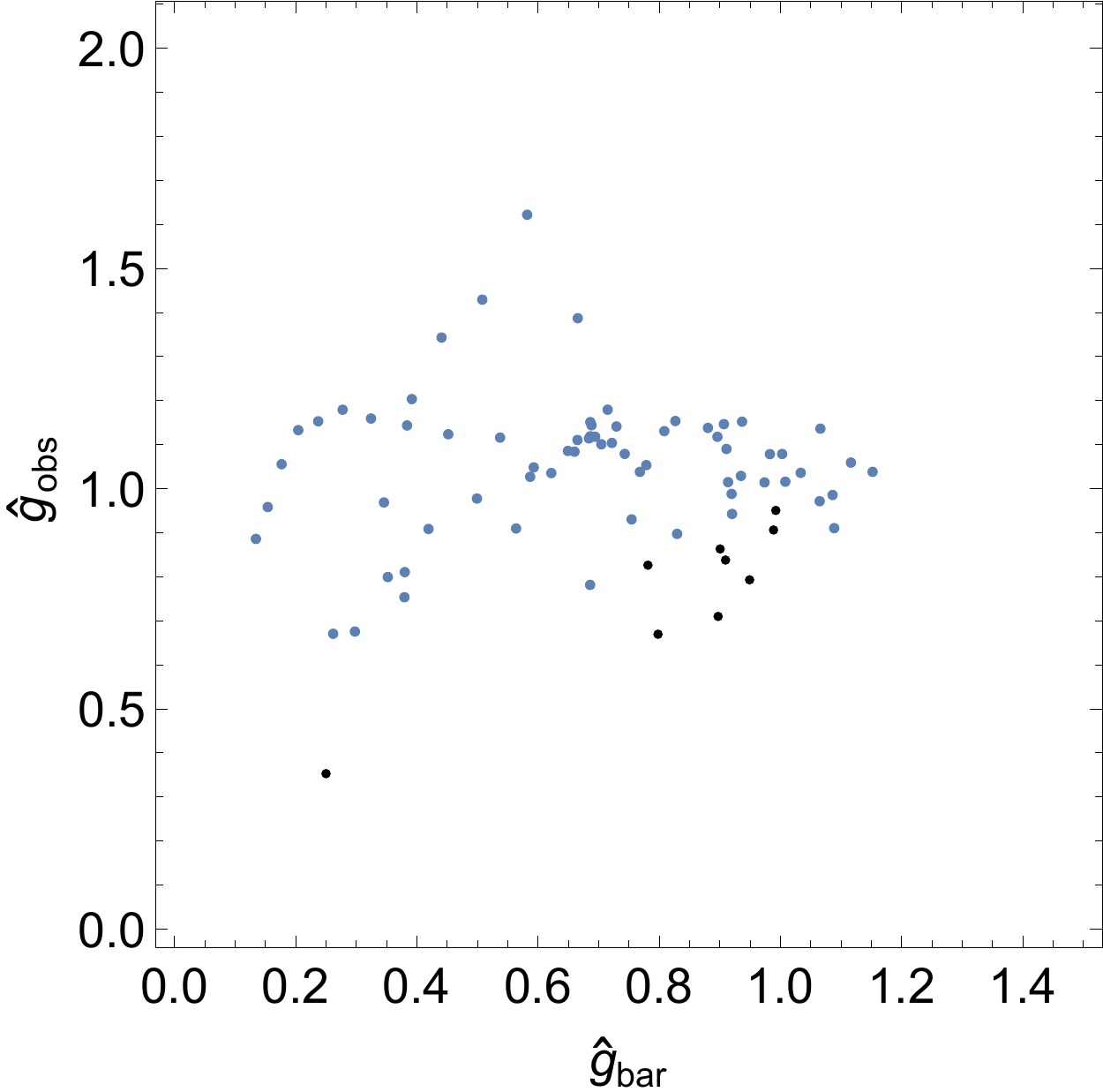}
	\end{subfigure}
	\begin{subfigure}{0.3\textwidth}
		\includegraphics[width=1\textwidth]{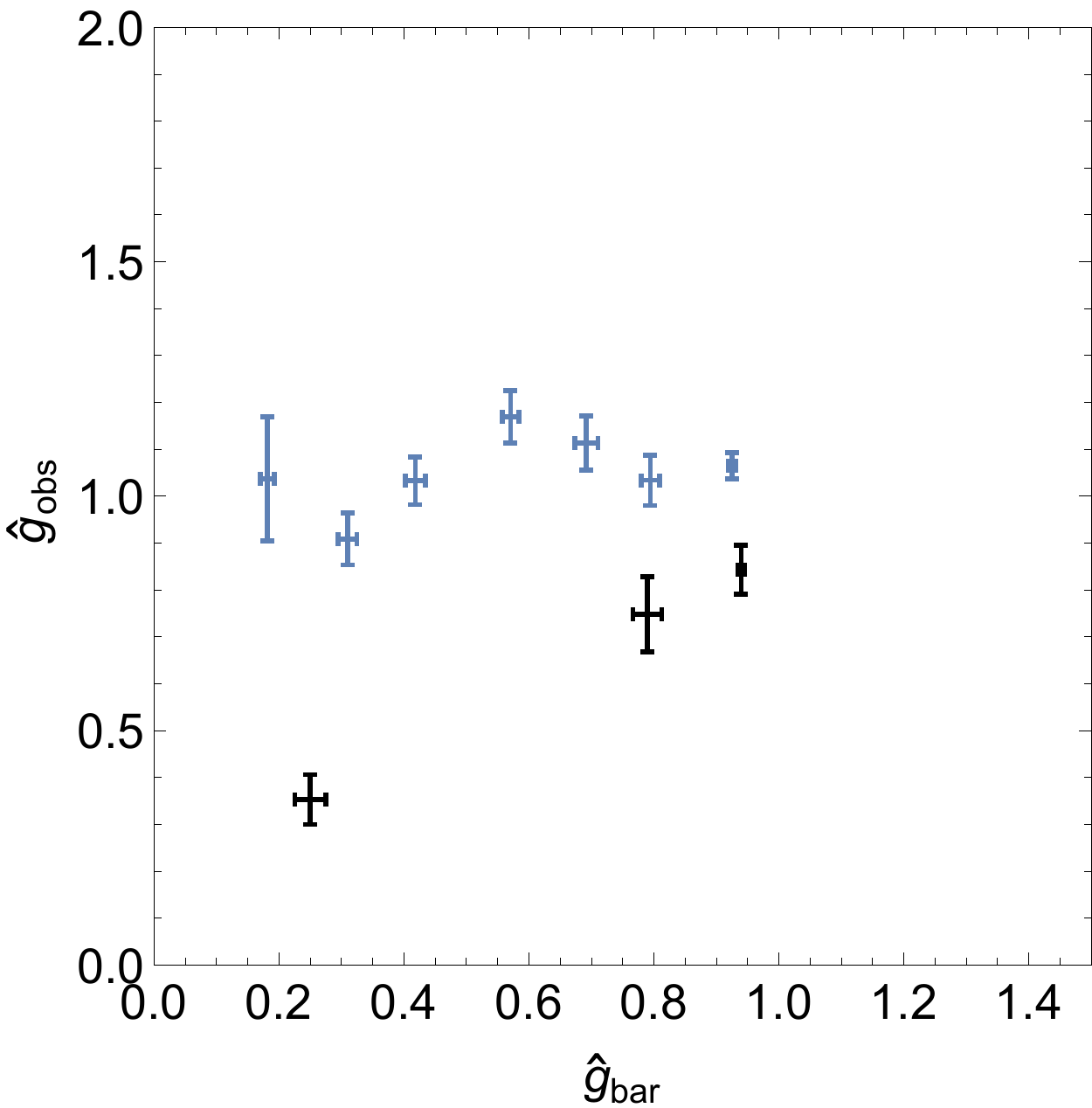}
	\end{subfigure}
	\caption
{Reproduction of figure \ref{fig:1} with $\frac{\delta v_{obs}(r_{j,i})}{v_{obs}(r_{j,i})}< 0.1$ imposed on individual data points.}
\label{fig:6}
\end{figure*}

In this appendix the impact of possible systematic uncertainties related to the relative magnitude of circular to non-circular motions in galaxies is discussed. The question here is whether or not these systematic uncertainties could be responsible for the systematic deviation between large and small radii that is observed. \citep{McGaugh:2016leg,Lelli:2017vgz} argue that discarding data points for which $\frac{\delta v_{obs}(r_{j,i})}{v_{obs}(r_{j,i})}\geq 0.1$ ensures that data points are not affected by strong non-circular motions. Figure \ref{fig:6} is a reproduction of figure \ref{fig:1} with $\frac{\delta v_{obs}(r_{j,i})}{v_{obs}(r_{j,i})}< 0.1$ imposed on individual data points. From the figure it is clear that although a significant amount of data points at small radii are removed, the overall tendencies remain unchanged.\newline\newline

\noindent Another approach is to consider the likelihood of the observed distribution of geometries (leftward$=2$, nowhere $=0$, rightward$=9$) begin caused by a systematic uncertainty that is random between galaxies. Due to the nowhere category being narrowly defined ($r_{bar}=r_{obs}$), I will be conservative and ignore this category, meaning that there would be a $50/50$ chance of a given galaxy turning out leftward or rightward. Given this information, the probability of observing $2$ leftward and $9$ rightward galaxies is $\lesssim 3\%$. Hence, it is unlikely to observe data under these circumstances.\newline \newline

\noindent In conclusion I find it unlikely that a systematic uncertainty within each galaxy is responsible for the deviation between large and small radii. I also find that removing points with large relative uncertainty on $v_{obs}$ does not change the overall tendencies. Coupled with the fact that systematic deviation between small and large radii is predicted by most solutions to the missing mass problem (e.g. NFW dark matter, pseudo-isothermal dark matter and MOND modified gravity), I do not find evidence to support a rejection of data points at small radii -- and the conclusions drawn therefrom in this article -- based on a possible systematic uncertainty related to the relative magnitude of circular to non-circular motions in galaxies.
\newpage

\end{appendices}

\bibliography{refs}

\end{document}